\documentclass[pra]{revtex4}
\usepackage[latin9]{inputenc}
\usepackage[dvipdfmx]{graphicx}
\usepackage{amsmath}
\usepackage{cases}
\usepackage{amssymb}
\usepackage{color}

\usepackage{amsfonts}
\newcommand{\ket}[1]{|{#1}\rangle}			
\newcommand{\bra}[1]{\langle{#1}|}			
		
\newcommand{\braket}[2]{\langle {#1}|{#2} \rangle}

\newcommand{\dev}{\mathrm{dev}}
\newcommand{\tr}{\mathrm{tr}}
\newcommand{\ad}{\mathrm{ad}}

\usepackage{amsthm}
\newtheorem{lemma}{Lemma}
\newtheorem{definition}{Definition}
\newtheorem{theorem}{Theorem}

\newtheorem{corollary}{Corollary}

\makeatother

\begin{document}

\title{
Lie-Algebraic Bounds on Quantum Control Time via the Baker-Campbell-Hausdorff Formula
}

\author{Go Kato$^1$}
\email{go.kato@nict.go.jp}
\author{Masaki Owari$^2$}
\author{Koji Maruyama$^{3,4}$}

\date{\today}

\affiliation{$^1$ Quantum ICT Laboratory, NICT, Koganei, Tokyo 184-8795, Japan\\
$^2$ Faculty of Informatics, Shizuoka University, Hamamatsu 432-8011, Japan \\
$^3$ Faculty of Engineering, University of Toyama, Toyama 930-8555, Japan \\
$^4$ Department of Chemistry and Materials Science, Osaka Metropolitan University, Osaka 558-8585, Japan
}

\begin{abstract}
The time required to implement a desired unitary operation is a central issue in quantum control, especially for many-body systems with limited control access. While controllability theory determines whether a target unitary is reachable in principle, it does not directly quantify the implementation time. Here we derive a Baker-Campbell-Hausdorff (BCH)-based inequality that connects these two questions at the operator level. Whenever a target unitary is implemented by the available Hamiltonians, an effective logarithmic generator can be chosen within the corresponding dynamical Lie algebra, and its normalized traceless Hilbert-Schmidt norm is bounded by the time integral of the same norm of the applied Hamiltonian. This induces a global-phase-insensitive distance on the reachable subgroup and yields a protocol-independent lower bound on the control time that explicitly respects Lie-algebraic restrictions. We compare the resulting bound with familiar quantum-speed-limit estimates and show that it refines a stabilizer-based control-time bound for an $XY$ spin chain in the single-excitation subspace. Our result thus provides an algebraic link between controllability and quantitative bounds on unitary implementation time.

\end{abstract}

\maketitle

\section{Introduction}

How long does it take to implement a desired unitary operation on a quantum system when only a limited set of Hamiltonians is experimentally accessible? This question is central to quantum control, because coherent operations must be completed within the finite coherence time of the physical system. It is also theoretically subtle. Standard Lie-algebraic controllability theory addresses the reachability question, namely whether a target unitary can be generated in principle, but it does not by itself quantify the control time required to synthesize it.

Quantum speed limits (QSLs) provide a natural framework for discussing lower bounds on evolution time; see, e.g., Refs.~\cite{Frey2016,DC17} for reviews. The Mandelstam-Tamm and Margolus-Levitin bounds~\cite{MT85,MargolusLevitin1998} relate the minimum time of quantum evolution to quantities such as the energy uncertainty or the mean energy. Later developments include geometric and operator-based formulations of QSLs and related bounds for unitary operations~\cite{Chau2011,Lee2012,Adolfo2021,Garcia2022,Poggi2019,Poggi2020,Hornedal2023}. Quantum brachistochrone and time-optimal-control formulations treat the implementation of a target state or unitary as a constrained optimization problem~\cite{Boscain2021,Carlini2006,Carlini2007,Koike2022,Yang2022,Khaneja2001,Khaneja2002,Caneva2009,Russell2015}. Shortcuts to adiabaticity and counterdiabatic driving provide another important perspective, constructing Hamiltonians that reproduce the outcome of an adiabatic process in finite time~\cite{Takahashi2013,Funo2017}.

The present work is related in spirit to these approaches, but its aim is different. We do not attempt to solve the full time-optimal control problem, nor do we construct an optimal or shortcut protocol. Instead, we derive a lower bound that any control protocol must satisfy if it implements a prescribed target unitary. The bound is formulated at the operator level and is compatible with the dynamical Lie algebra generated by the available Hamiltonians.

In a typical closed-system control setting, the Hamiltonian takes the form
\begin{equation}
H(t)=H_0+\sum_m h_m(t)H_m,
\end{equation}
where $H_0$ is a drift Hamiltonian and $H_m$ are directly controllable Hamiltonians. The corresponding dynamical Lie algebra generated by $\{iH_0,iH_1,...,iH_M\}$ determines the reachable unitary group. This Lie-algebraic controllability criterion is powerful, but it is essentially a reachability statement. It does not directly give a lower bound on the time required to implement a specific unitary.

This difficulty is closely tied to the noncommutativity of the available Hamiltonians. Although only a small set of Hamiltonians may be directly accessible, their commutators can generate additional directions and may span a large dynamical Lie algebra~\cite{dalessandroBook,Lloyd2004,Schirmer2008,bmm2010,bmBookChapter}. Algebraic
structures induced by limited direct access have also been investigated in related settings~\cite{kom2019}. Converting such local algebraic information into global lower bounds on the time required for a specified target unitary remains challenging. Related time-optimal analyses have been carried out in special low-dimensional
settings~\cite{boozer2012}.

Our main result is a Baker-Campbell-Hausdorff (BCH)-based inequality for effective generators of controlled unitary evolutions. 
Roughly speaking, if a target unitary is implemented by a time-dependent Hamiltonian generated by the available controls, then one can choose an effective generator of the target inside the corresponding dynamical Lie algebra. After the identity components are removed, the size of this generator is bounded by the time integral of the size of the applied Hamiltonian. This immediately gives a lower bound of the form
\begin{equation}
\mathrm{time}
\geq
\frac{\mathrm{distance\ to\ the\ target\ in\ the\ reachable\ algebra}}
{\mathrm{maximum\ available\ speed}}.
\end{equation}
In this sense, the bound is projective: it removes the physically irrelevant global phase of the resulting unitary. This is achieved by using a normalized traceless Hilbert-Schmidt norm, denoted below by $\operatorname{dev}$. In this way, the result provides a direct algebraic connection between controllability and quantitative lower bounds on unitary implementation time.

The BCH formula is relevant because a time-dependent unitary evolution is naturally approximated by a product of short-time exponentials. Since the usual BCH series is guaranteed to converge only near the identity, we apply it to sufficiently small pieces and then take an appropriate limit. The result is an inequality of the form $\dev C \leq \dev A+\dev B$ for an effective generator $C$ satisfying $e^C=e^Ae^B$, which can then be iterated to derive lower bounds on control time.

The scope of this result should be stated carefully. The distance to the target is defined by minimizing not over the full set of Hermitian logarithmic generators $C$ satisfying $e^{-iC}=U^\mathrm{(target)}$, but only over those satisfying the Lie-algebra constraint $iC\in\mathcal L$. The bound therefore respects the algebraic constraint $iC\in\mathcal L$ on the effective logarithmic generator, but it does not distinguish directions generated directly by available Hamiltonians from those arising only through high-order commutators. When the dynamical Lie algebra is the full $\mathfrak{su}(D)$, the distance appearing in the bound reduces essentially to a standard projective-unitary logarithmic distance. The present result should therefore be regarded as a first algebraic lower bound, rather than a complete solution of the many-body time-optimal control problem.

This point also clarifies the relation to quantum brachistochrone and shortcut-to-adiabaticity approaches. Those approaches typically seek an optimal or accelerated Hamiltonian protocol under given constraints. By contrast, the present work gives a necessary condition that any such protocol must obey, once it is described within the dynamical Lie algebra. In this sense, the result is complementary to, rather than a replacement for, optimal-control or shortcut-construction methods.

We compare the bound with familiar QSL-type estimates using the Choi representation. The purpose is not to claim a universal improvement over QSLs, but to clarify the difference between a state-space distance and a unitary-level projective distance. 
We also illustrate the bound using the XY spin-chain example of Lee et al. in the single-excitation subspace~\cite{Lee2018}, where the proposed distance refines a stabilizer-based control-time bound.

The paper is organized as follows. In Sec.~\ref{sec:main_result}, we formulate the control setting, introduce the deviation norm, and state the main BCH-based inequality and the induced projective distance. In Sec.~\ref{sec:comparison}, we compare the resulting bound with state-based and unitary-based speed-limit estimates. In Sec.~\ref{sec:OutlineOfProof}, we outline the proof based on the BCH formula. In Sec.~\ref{sec:conclusion}, we discuss the geometric interpretation of the Hilbert-Schmidt-norm version of the result, summarize the limitations of the present bound, and comment on future directions.

\section{Main result}
\label{sec:main_result}

\subsection{Control setting and dynamical Lie algebra}

We consider dynamics governed by the Schr\"odinger equation
\begin{equation}
i\frac{d}{dt}U(t)=H(t)U(t),
\label{eq:schrodinger_operator}
\end{equation}
with the initial condition
\begin{equation}
U(0)=I.
\label{eq:initial_condition}
\end{equation}
Here $U(t)$ is a unitary operator, $H(t)$ is a time-dependent Hamiltonian, and we set $\hbar=1$.

The control Hamiltonian is assumed to have the form
\begin{equation}
H(t)=H_0+\sum_{m=1}^{M} h_m(t)H_m,
\label{eq:control_hamiltonian}
\end{equation}
where $H_0$ is a drift Hamiltonian and $H_m$ are directly controllable Hamiltonians. The real-valued functions $h_m(t)$ represent the applied control fields. We assume that $H(t)$ is piecewise continuous in time. 

With amplitude constraints on the control fields, the allowed instantaneous Hamiltonians form
\begin{equation}
\mathcal{H}_{\rm exp} = \left\{ H=H_0+\sum_{m=1}^{M} h_m H_m \ \middle|\  |h_m|\leq h_m^{\max} \right\}.
\label{eq:hexp}
\end{equation}
We denote by
\begin{equation}
\mathcal{L} = \operatorname{Lie}\{iH_0,iH_1,\ldots,iH_M\}
\label{eq:dla_definition}
\end{equation}
the dynamical Lie algebra generated by the drift and control Hamiltonians, namely the Lie closure under repeated commutators and their linear combinations.

\subsection{Deviation norm and removal of global phase}

For a $D\times D$ operator $A$, define the deviation norm of an operator by
\begin{equation}
\left(\operatorname{dev}A\right)^2 :=
\frac{1}{D}\operatorname{tr}(A^\dagger A) - \left| \frac{1}{D}\operatorname{tr}A \right|^2 .
\label{eq:dev_definition}
\end{equation}
Equivalently,
\begin{equation}
\operatorname{dev}A = \frac{1}{\sqrt D} \left\| A-\frac{\operatorname{tr}A}{D}I \right\|_F ,
\label{eq:dev_traceless_frobenius}
\end{equation}
where $\|\cdot\|_F$ is the Frobenius norm, $\left\|A\right\|_F:=\sqrt{\mathrm{tr} A^\dagger A}$, which is often called the Hilbert-Schmidt norm in the literature.
This dev is a legitimate norm for traceless operators, while it is a semi-norm otherwise.
Thus $\operatorname{dev}A$ measures the size of $A$ after removing its identity component. For Hermitian or anti-Hermitian normal operators, it is the standard deviation of the eigenvalues, up to the corresponding factor of $i$ in the anti-Hermitian case.

This definition is independent of any particular quantum state. We use the notation $\operatorname{dev}$ to distinguish it from the usual state-dependent uncertainty of an observable
\begin{equation}
(\Delta A)^2 = \langle \phi|A^2|\phi\rangle - \langle \phi|A|\phi\rangle^2 .
\label{eq:state_dependent_uncertainty}
\end{equation}
Since the trace component is removed, the distance induced by $\operatorname{dev}$ naturally lives on a projective unitary group rather than on the full unitary group.

\subsection{BCH inequality for two operators}

The key algebraic ingredient is the following lemma.

\begin{lemma}
\label{lem:bch_dev_triangle}
Let $A$ and $B$ be anti-Hermitian operators. Then there exists an anti-Hermitian operator $C$ such that
\begin{equation}
e^C=e^A e^B,
\label{eq:lemma_product}
\end{equation}
\begin{equation}
C\in\operatorname{Lie}\{A,B\},
\label{eq:lemma_lie_membership}
\end{equation}
and
\begin{equation}
\operatorname{dev}C \leq \operatorname{dev}A+\operatorname{dev}B,
\label{eq:lemma_dev_triangle}
\end{equation}
\begin{equation}
\operatorname{tr}C = \operatorname{tr}(A+B).
\label{eq:lemma_trace}
\end{equation}
\end{lemma}

At first sight, Eq.~\eqref{eq:lemma_product} resembles the BCH formula. The point of the lemma, however, is not merely the formal identity $e^Ae^B=e^C$. The usual BCH series is guaranteed to converge only when $A$ and $B$ are sufficiently small.
Lemma~\ref{lem:bch_dev_triangle} asserts the existence of a suitable $C$ in the Lie closure $\mathrm{Lie}\{A,B\}$, together with the deviation inequality~\eqref{eq:lemma_dev_triangle}, without assuming that the original $A$ and $B$ are small. The proof is given in Sec.~\ref{sec:OutlineOfProof} and in Sec.~VIII of the Supplemental Material. The anti-Hermitian assumption is essential for the argument; Sec.~I of the Supplemental Material gives a simple example showing that the analogue of Eq.~\eqref{eq:lemma_lie_membership} need not hold for non-anti-Hermitian operators.

\subsection{Main theorem}

We now state the main result for a continuously modulated Hamiltonian.

\begin{theorem}
\label{thm:main_dev_bound}
Suppose that a target unitary $U^{(\mathrm{target})}$ is implemented at time $T$ by a solution of Eqs.~\eqref{eq:schrodinger_operator} and \eqref{eq:initial_condition}, with $H(t)$ of the form Eq.~\eqref{eq:control_hamiltonian}. Then there exists a Hermitian operator $C[T]$ such that
\begin{equation}
e^{-iC[T]}=U^{(\mathrm{target})},
\label{eq:theorem_target_generator}
\end{equation}
\begin{equation}
iC[T]\in\mathcal{L}
\label{eq:theorem_lie_membership}
\end{equation}
and
\begin{equation}
\operatorname{dev}C[T] \leq \int_0^T \operatorname{dev}H(t)\,dt,
\label{eq:theorem_dev_bound}
\end{equation}
\begin{equation}
\operatorname{tr}C[T] = \int_0^T \operatorname{tr}H(t)\,dt.
\label{eq:theorem_trace}
\end{equation}
\end{theorem}

The notation $C[T]$ does not imply that $C$ is a single-valued continuous function of $T$. It indicates that $C[T]$ is an effective generator associated with a particular control protocol that reaches the target at time $T$. For a fixed target unitary, there may be multiple protocols and hence multiple possible effective generators. The possible discontinuity of such a choice of $C[T]$ is illustrated by a simple example in Sec.~II of the Supplemental Material.

Theorem~\ref{thm:main_dev_bound} is obtained by approximating the time-ordered evolution by a product of short-time exponentials and applying Lemma~\ref{lem:bch_dev_triangle} recursively. Taking the continuum limit gives Eqs.~\eqref{eq:theorem_target_generator}--\eqref{eq:theorem_trace}. In this sense, the theorem is an operator-level consequence of the BCH structure of controlled unitary evolution.

\subsection{Lower bound on control time}

Theorem~\ref{thm:main_dev_bound} immediately implies a lower bound on the implementation time. Define
\begin{equation}
c_{\mathcal{L}}\!\left(U^{(\mathrm{target})}\right) := \min
\left\{
\operatorname{dev}C \ \middle|\ e^{-iC}=U^{(\mathrm{target})}, \ iC\in\mathcal{L}
\right\}.
\label{eq:cL_definition}
\end{equation}
Then Eq.~\eqref{eq:theorem_dev_bound} gives
\begin{equation}
c_{\mathcal{L}}\!\left(U^{(\mathrm{target})}\right) \leq \int_0^T \operatorname{dev}H(t)\,dt \leq T\max_{0\leq t\leq T}\operatorname{dev}H(t),
\label{eq:cL_integral_bound}
\end{equation}
and therefore
\begin{equation}
T \geq \frac{c_{\mathcal{L}}\!\left(U^{(\mathrm{target})}\right)}{\max_{0\leq t\leq T}\operatorname{dev}H(t)}.
\label{eq:time_lower_bound}
\end{equation}
This has the form ``distance divided by maximum speed.'' The numerator is the minimum projective generator size of the target within the reachable Lie algebra. The denominator is the maximum projective Hamiltonian strength available along the implemented protocol. We denote the right-hand side of Eq.~(\ref{eq:time_lower_bound}) by $T_*$.

In the special case where $\mathcal{L}=\mathfrak{su}(D)$, the minimization in Eq.~\eqref{eq:cL_definition} reduces to the usual minimization over logarithm branches, modulo global phase:
\begin{equation}
c_{\mathfrak{su}(D)}(U)
=
\min_{\theta\in\mathbb{R}}
\operatorname{dev}
\left[
i\,\operatorname{Log}\left(Ue^{i\theta}\right)
\right],
\label{eq:csuD}
\end{equation}
where $\operatorname{Log}$ denotes the principal logarithm. In this case the bound is essentially a projective-unitary logarithmic distance divided by the available speed. For a proper subalgebra or a nontrivial representation of the reachable algebra, the value of $c_{\mathcal{L}}$ can differ from the full-$\mathfrak{su}(D)$ value.

Equation~\eqref{eq:time_lower_bound} should be interpreted as a lower bound obtained under the algebraic constraint $iC\in\mathcal L$. It does not, however, estimate the time overhead required to synthesize the minimizing generator $C$ from the available Hamiltonians $\{H_0, H_1,..., H_M\}$. Directions that arise only through high-order commutators may therefore require additional control time beyond what is captured by this bound.

The bound is nevertheless saturable in simple cases. Suppose that the Hamiltonian is always proportional to a fixed traceless Hermitian operator $H_\ast$ with $iH_\ast\in\mathcal{L}$:
\begin{equation}
H(t)=f(t)H_* .
\label{eq:single_direction_hamiltonian}
\end{equation}
For the target
\begin{equation}
U^{(\mathrm{target})}=e^{-igH_\ast},
\label{eq:single_direction_target}
\end{equation}
and when the logarithmic branch $C=gH_\ast$ gives $c_{\mathcal L}=|g|\operatorname{dev}H_\ast$, the minimum time under the amplitude bound $|f(t)|\leq f_{\max}$ is
\begin{equation}
T_{\min} = \frac{|g|}{f_{\max}} = \frac{c_{\mathcal{L}}}{f_{\max}\operatorname{dev}H_\ast}.
\label{eq:saturation_example}
\end{equation}
This example shows that the bound can be saturated in simple constant-Hamiltonian cases. It should not be read, however, as a statement that the bound is saturated for generic control systems.

\subsection{Projective distance induced by the theorem}

Theorem~\ref{thm:main_dev_bound} also induces a distance on the reachable subgroup modulo global phases. For two unitaries $U_A$ and $U_B$ in the reachable group, define
\begin{equation}
\widetilde d(U_A,U_B) := \min
\left\{ \operatorname{dev}C
\ \middle|\ 
e^{-iC}=U_AU_B^{-1},\ iC\in\mathcal{L} \right\}.
\label{eq:projective_distance}
\end{equation}
Because $\operatorname{dev}C=0$ precisely when $C$ is proportional to the identity, this distance identifies unitaries that differ only by a global phase. Lemma~\ref{lem:bch_dev_triangle} implies the triangle inequality,
\begin{equation}
\widetilde d(U_A,U_B)
\leq
\widetilde d(U_A,U_C)
+
\widetilde d(U_C,U_B),
\label{eq:triangle_distance}
\end{equation}
and therefore $\widetilde d$ defines a metric on the corresponding subgroup of the projective unitary group, $PU(D)=U(D)/U(1)$, whenever the relevant minimizers exist within the reachable algebra.

With this notation, Eq.~\eqref{eq:cL_definition} can be written as
\begin{equation}
c_{\mathcal{L}}\!\left(U^{(\mathrm{target})}\right)
=
\widetilde d\!\left(U^{(\mathrm{target})},I\right),
\label{eq:cL_as_distance}
\end{equation}
and the time bound becomes
\begin{equation}
T
\geq
\frac{
\widetilde d\!\left(U^{(\mathrm{target})},I\right)
}{
\max_{0\leq t\leq T}\operatorname{dev}H(t)
}.
\label{eq:time_bound_distance}
\end{equation}

Because the minimization in Eq.~\eqref{eq:projective_distance} is constrained by the Lie algebra $\mathcal L$, the resulting distance can depend on the chosen dynamical Lie algebra and its representation, even for the same pair of unitaries. A specific example illustrating this algebra dependence is given in Sec.~III of the Supplemental Material. When $\mathcal L$ is the full $\mathfrak{su}(D)$, the distance reduces to the usual projective logarithmic distance. When $\mathcal L$ is a proper subalgebra, the minimization is constrained by the allowed effective generators, and the resulting distance can be larger or may be undefined for unreachable targets.

The next section compares this projective operator-level bound with speed-limit estimates based on state-space distances and with related unitary-distance bounds.

\section{Comparison with quantum speed limits and unitary-distance bounds}\label{sec:comparison}

\subsection{The Mandelstam-Tamm bound}
We now compare our bound, denoted by $T_*$, with those implied by well-known discussions on quantum speed limits. Although the bound obtained by Margolus and Levitin (ML) is a major speed limit for time-independent Hamiltonians, there is no widely accepted extension to time-dependent closed-system dynamics \cite{DeffnerPRL2013,DeffnerJPA2013,Okuyama2018a,Okuyama2018b}.
In fact, H\"{o}rnedal and S\"{o}nnerborn constructed explicit counterexamples showing that an ML-type estimate based on the time-averaged normalized mean energy can be violated \cite{Hoernedal2023}.
Thus, we focus on the speed limit by Mandelstam and Tamm (MT) \cite{MT85}, whose original derivation was based on the uncertainty relationship between energy and time. 

There is a subtle difference in the formulations of their bounds and ours: While the MT bound measures the speed of evolution in terms of the \textit{fidelity} between the initial and final (target) states, we consider the time to obtain the target unitary operation $U^\mathrm{(target)}$ since our main interest is in the implementation of quantum control operations rather than reaching a specific fidelity with respect to the initial state. 
Although the two bounds are formulated for different tasks, it is possible to connect these measures through the Choi representation of quantum operations; i.e., in the operator-form comparison considered below, our projective-distance bound can be sharper than the MT-type state-distance bound.

The setting for the MT bound is similar to the one given by Eqs. (\ref{eq:schrodinger_operator}) and (\ref{eq:initial_condition}). A quantum state is in $\ket{\phi(0)}$ at $t=0$, and it evolves according to the Schr\"odinger equation. Letting $\ket{\phi(\tau)}$ be the target state, it is summarized as follows. 

\textit{Mandelstam-Tamm bound:}  The fidelity, 
$|\braket{\phi(0)}{\phi(\tau)}|^2$, or the Bures angle \cite{Jozsa1994}, is bounded by 
\begin{equation}\label{MTbound00}
\arccos |\braket{\phi(0)}{\phi(\tau)}| \leq \int _{0}^\tau dt \Delta H(t),
\end{equation}
where $(\Delta A)^2=\bra{\phi (t)}A^2\ket{\phi(t)}-\left(\bra{\phi (t)}A\ket{\phi(t)}\right)^2$ is the deviation of operator $A$. The MT bound is often written as the time needed for a state to become orthogonal to the initial state, i.e., $\tau \ge \pi/(2\Delta H_\tau)$, with $\Delta H_t=(1/t)\int_0^t \Delta H(t') dt'$.
A closely related bound, where $\Delta H$ is replaced with the time-averaged norm of $H(t)$, has also been derived without referring to specific states. In \cite{Aifer2022}, a lower bound for the transition time of the (known) geometric angle between the (unknown) initial and final states is obtained by considering the minimum energy cost. 

The MT bound can also be seen as a consequence of geometrical considerations a la Anandan and Aharanov \cite{Anandan1990}, as well as Uhlmann \cite{Uhlmann1992}. The length of the path $\mathcal{C}$ that a quantum state follows under the Schr\"{o}dinger equation was derived in \cite{Anandan1990} as
\begin{equation}\label{AnandanLength}
\mathrm{length}(\mathcal{C}) = 2\int_0^T \Delta H(t) dt,
\end{equation}
where the length is evaluated with the Fubini-Study metric \cite{Bengtsson2017}. 
In a similar spirit to Eq. (\ref{AnandanLength}), Poggi has obtained inequalities for quantum speed limits for unitary transformations \cite{Poggi2019}, which we look at in the next subsection.

We compare Eq. (\ref{MTbound00}) with Eq. 
%(\ref{eq:main_5})
(\ref{eq:theorem_dev_bound}) by rewriting them in terms of operators through the Choi representation. To do this, suppose that the Hilbert space in which the quantum states reside is a tensor product of two $D$-dimensional spaces, i.e., $\mathcal{H}^{\otimes 2}$. The Hamiltonian $H(t)$ is replaced with $I\otimes H(t)$, and the initial state is set to be
$\ket{\phi(0)}=1/\sqrt{D} \sum_n \ket{n}\ket{n}$. Then, we have the following corollary: 

% Corollary 1
\begin{corollary}\label{cor_choi_MT}
(Operator form of the Mandelstam-Tamm bound) Let the unitary operator $U(T)$ be a solution of Eq. (\ref{eq:schrodinger_operator}) evaluated at time $T$, with the initial state $\ket{\phi(0)}=1/\sqrt{D} \sum_n \ket{n}\ket{n}$. The MT bound, Eq. (\ref{MTbound00}), is rewritten as
\begin{equation}\label{MTbound_choi1}
\arccos\left| D^{-1} \mathrm{tr} U(T) \right| \leq \int_0^T \dev H(t) dt,
\end{equation}
where $(\dev A)^2:=D^{-1}\mathrm{tr}(A^2)-(D^{-1}\mathrm{tr}A)^2$. 
\end{corollary}

\subsection{Comparison with a specific example}
\label{sec:comparisonMT}
Letting $H(t)=C[T]/T$ in Eq. (\ref{MTbound_choi1}) of Corollary \ref{cor_choi_MT} so that $U(T)=e^{-iC[T]}=U^\mathrm{(target)}$, we obtain
\begin{equation}\label{comparisonMT}
\arccos\left|D^{-1}\mathrm{tr} U^\mathrm{(target)}\right| \le \dev C[T].
\end{equation}
The LHS of Eq. (\ref{comparisonMT}) is the lower MT bound from Eq. (\ref{MTbound_choi1}), and its RHS is the lower bound from our Theorem \ref{thm:main_dev_bound}. 
Therefore, in this operator-form comparison, Eq.~(\ref{comparisonMT}) shows that our bound is no smaller than the MT-type bound.
Note that, although there may exist a $C\notin \mathcal{L}$ such that $e^{-iC}=U^\mathrm{(target)}$ for a given target unitary, Eq. (\ref{comparisonMT}) still holds.

To see the difference between the bounds of Theorem \ref{thm:main_dev_bound} and Corollary \ref{cor_choi_MT} more clearly, let us consider a specific example in which the Hamiltonian $H(t)$ is given as the following, somewhat artificial, function of time for $m\in \mathbb{N}$:
\begin{align}\label{zigzagH}
 H(t) &:=\left\{
 \begin{array}{cl}
 A, &\makebox{when } 2m\leq t < 2m+1
 \\
 B, &\makebox{when } 2m+1\leq t < 2m+2
 \end{array}
 \right.
 \end{align}
 where 
 \begin{align}
 \begin{split}
 A &=
 \left( \begin{array}{ccc}
 a & b e^{i\theta} & 0 \\
 b e^{-i\theta} & -a & 0 \\
 0 & 0 & c
 \end{array}
 \right),
 \\
 B &=\left(
 \begin{array}{ccc}
 a & -b e^{-i\theta} & 0 \\
 -b e^{i\theta} & -a & 0 \\
 0 & 0 & c
 \end{array} \right),
 \end{split}
 \\
 \theta &= \arctan \left(\frac{a}{\sqrt{a^2+b^2}}\tan\sqrt{a^2+b^2}\right).
\end{align}
The parameters are set as $a>0$, $\sqrt{a^2+b^2}<\pi/2$, and $0\leq c<\theta< \pi/2$. By letting $b$ be small compared with $a$, sequential applications of $A$ and $B$ will eventually lead to a zigzag path along the ``straight" path generated by $\mathrm{diag}(\theta,-\theta,c)$. 

We now compare the time needed to implement unitary operations, $U(t)$, generated by the Hamiltonian in Eq. (\ref{zigzagH}) when $t=2M\; (M\in \mathbb{N})$. The corresponding unitary matrix takes the simple form 
 \begin{align}\label{u2m}
 U(2M) &= (e^{-iB}e^{-iA})^M
 \nonumber\\
 &\equiv
 \left(
 \begin{array}{ccc}
 e^{-2 i \theta M} & 0 & 0 \\
 0 & e^{2 i\theta M} & 0 \\
 0 & 0 & e^{- 2 ic M}
 \end{array} \right),
 \end{align}
where the rightmost matrix is the representation after diagonalization. The Hamiltonian is obviously in $\mathcal{L}({i A, i B})$ for all $t$, and the real time needed to realize $U(2M)$ in Eq. (\ref{u2m}) is $2M$ (for $cM \le \pi$).

Noting that $(\mathrm{dev}H(t))^2=2/3 (a^2+b^2+c^2/3)$ is a constant, the MT bound $T_{MT}$ is obtained from Eq. (\ref{MTbound_choi1}) as
\begin{align}\label{ex_tMT}
T_{MT} &= \frac{1}{\mathrm{dev}H(t)}\arccos\left| D^{-1} \mathrm{tr} U(2M) \right| \nonumber \\
&= \sqrt{\frac{3}{2}}\frac{1}{\sqrt{a^2+b^2+c^2/3}} \arccos \left| \frac{2\cos 2\theta M +e^{-2ic M}}{3}\right|.
\end{align}

The bound $T_*$ yielded by our result, Eq. (\ref{eq:theorem_dev_bound}), can be written as
\begin{align}\label{ex_tKOM}
T_* &= \frac{\dev C[2M]}{\dev H(t)} \nonumber \\
&= \frac{2}{\sqrt{a^2+b^2+c^2/3}}\min_{j,k\in\mathbb N} 
\sqrt{\left(\theta M-j \pi\right)^2+\frac 13\left(c M-k \pi \right)^2},
\end{align}
where $C[2M]=i \log U(2M)$. This $C[2M]$ lies in $\mathcal{L}(\{iA,iB\})$, which is strictly smaller than $\mathfrak{su}(3)$. If it were taken from the full $\mathfrak{su}(3)$, the bound $T_*$ would be smaller than that in Eq. (\ref{ex_tKOM}). This is a consequence of considering the algebraic structure. 

Figure 1 shows a comparison between $T_{MT}$ and $T_*$ in which $a=\pi/500, b=-\pi/1200, c=\pi/1000$ are set, as well as a comparison 
with the real control time $T_\mathrm{real}=2M$ needed to realize the unitary operator in Eq. (\ref{u2m}). For this example, our bound is sharper than the MT-type bound and is closer to the real control time, $2M$.

\begin{figure}
\includegraphics[scale=0.6]{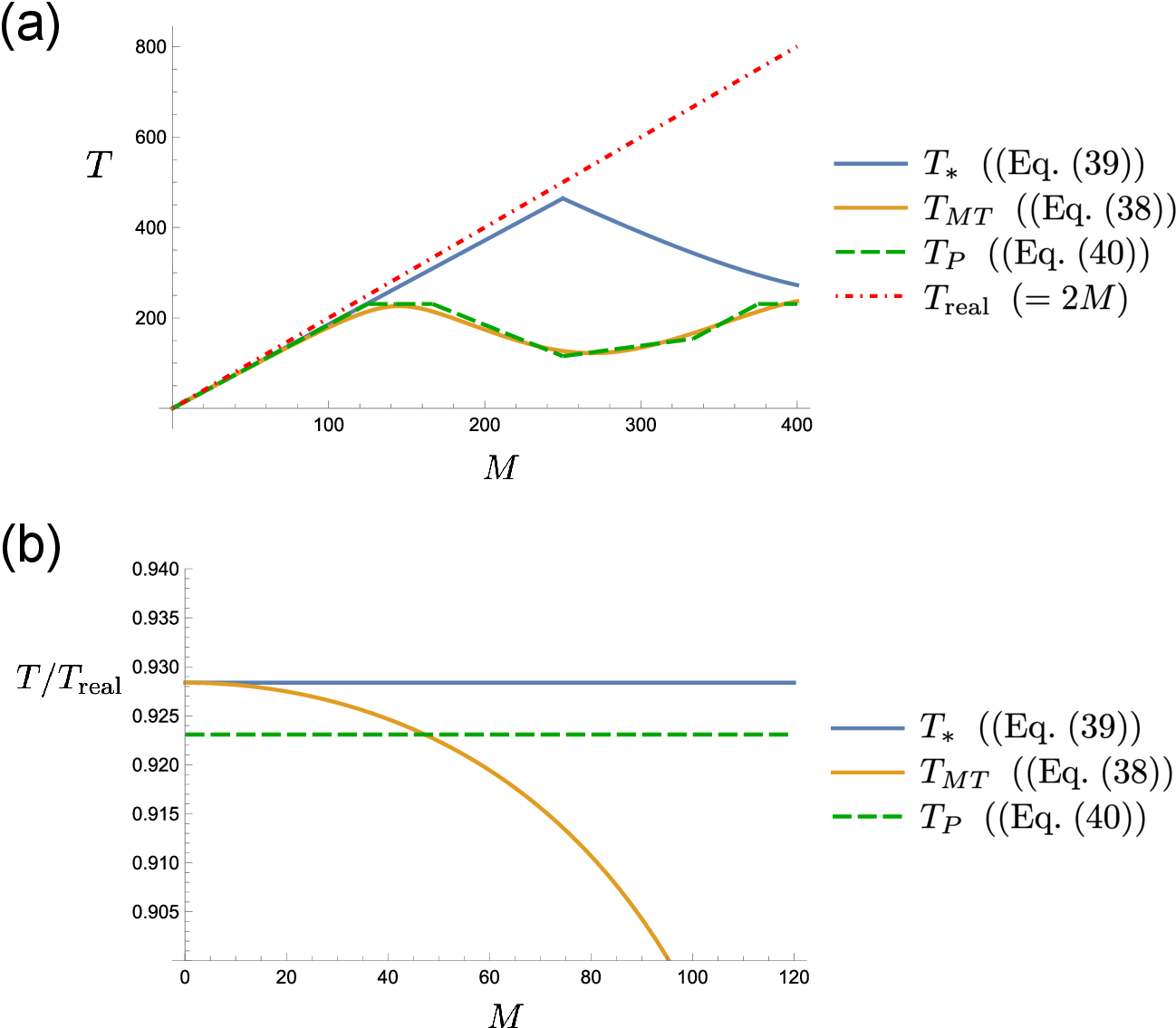}
\caption{
(a) Comparison between the control times needed to implement $U(2M)$ in Eq. (\ref{u2m}) generated by the Hamiltonians in Eq. (\ref{zigzagH}). The red dot--dashed line indicates the real achievable control time. Owing to the cyclic nature of $U(2M)$ as well as nonanalytic operations in the expressions of the control times, these times have curved or angular profiles. $T_\mathrm{real}$ is a straight line, simply because it is based on Eq. (\ref{u2m}). (b) Ratios of the control times to $T_\mathrm{real}$ for a short time starting at 
$t=0\; (M=0)$. The flatness of the lines indicates that the corresponding estimate considers the algebraic structure correctly. }
\label{fig:comparison}
\end{figure}

In some parameter regimes, $T_*$ gives a substantially larger lower bound than $T_{\rm MT}$, for example in the vicinity of $M\simeq 250$ in Fig.~1(a). This is because $T_*$ is defined using an effective generator restricted to the dynamical Lie algebra generated by $A$ and $B$. By contrast, the MT-type estimate, and similarly Poggi's bound discussed below, do not encode this algebraic constraint and may therefore allow algebraically inadmissible shortcuts, leading to shorter lower-bound estimates.

The blue curve for $T_*$ in Fig. 1(a) begins to decrease at $M\simeq 250$ because of the discontinuity when evaluating the principal value of $\log U$ and the minimization in Eq. (\ref{ex_tKOM}). Also, $T_{MT}$ and $T_P$ reach their peaks at $T\simeq 150$ since the state becomes orthogonal to the initial state; roughly speaking, the state reaches the opposite point on the great circle on a sphere and starts to return.
Figure~1(b) compares the bounds near the identity operation, where the ratios to the real time $T_{\rm real}$ are plotted in order to observe the intrinsic difference due to the way in which distances are measured. In this near-identity regime, the comparison highlights the effect of using an operator distance restricted to the dynamical Lie algebra rather than a state-space distance.

As noted above in Sec. \ref{sec:comparison}A, Poggi derived quantum speed limits for unitary transformations in \cite{Poggi2019}, following a geometrical approach. These limits are given in Eqs. (12) and (16) in \cite{Poggi2019}, and here, we compare them with that in Eq. (12), since it provides the sharper of the two bounds for the present comparison. It can be evaluated for the present example of Eq. (\ref{zigzagH}) as 
\begin{align}
T_P
&=\frac{2\arccos{\min_{\psi}\left|\bra{\psi}U(2M)\ket{\psi}\right|}}{E_\mathrm{max}(t)-E_\mathrm{min}(t)}
\nonumber\\
&= \frac{\min\left(\phi,\pi\right)}{2\sqrt{a^2+b^2}},
\label{TPoggi2019}
\end{align}
where $E_\mathrm{max}(t)$ and $E_\mathrm{min}(t)$ are the largest and smallest eigenvalues of $H(t)$, respectively, and 
\begin{align}\label{TPoggi2019phi}
\phi:=2\pi-2\times \left\{
\begin{array}{ll}
\max(\pi-2\theta M,(\theta+c)M), & \makebox{if $0\leq 2\theta M\leq \pi$}\\
\max(2\theta M-\pi,\pi-(\theta-c)M), & \makebox{if $\pi < 2\theta M$ and $2(\theta +c)M\leq 2\pi$}\\
\max(\pi-2\theta M,(\theta+c)M-\pi,(\theta-c)M), & \makebox{if $2\pi<2(\theta +c)M$ and $\pi < 2\theta M\leq 2\pi$}\\
\end{array}
\right.
\end{align}

The nonanalyticity of $T_P$, due to the minimum and maximum operations in Eqs. (\ref{TPoggi2019}) and (\ref{TPoggi2019phi}), is shown in the plot of $T_P$ (green dashed line) in Fig. 1. The derivation of $T_P$ is based on the `fidelity' between the initial and target states; thus, its overall behavior is similar to that of the MT bound.

\subsection{Relation to other speed limits}
In order to compare our result with other speed limits, let us first prove the following lemma from Lemma \ref{lem:bch_dev_triangle}.
\begin{lemma}
\label{th:lemma_1}
For two anti-Hermitian operators, $A$ and $B$, there is an anti-Hermitian $C$ such that 
\begin{align}
\label{eq:lemma1_cond1}
e^{C} &= e^{A}e^{B},
\\
\label{eq:lemma1_cond2}
C &\in \mathcal L(\{A,B\}),
\\
\label{eq:lemma1_cond3}
\mathrm{and} \;\;\|C\|_F &\leq \|A\|_F+\|B\|_F.
\end{align}
\end{lemma}

%This lemma can be proven as follows.

\begin{proof}
For any $D\times D$ anti-Hermitian matrix $X$, define the traceless matrix $\tilde X:=X-(\tr X/D)I$. Then
\begin{equation}
\|X\|_F^2=\|\tilde X\|_F^2+\frac{|\tr X|^2}{D}.
\end{equation}
By Lemma~\ref{lem:bch_dev_triangle}, there exists $C\in\mathcal L(\{A,B\})$ such that
$e^C=e^Ae^B$, $\dev C\le \dev A+\dev B$, and $\tr C=\tr(A+B)$.
Since $\dev \tilde{X}=D^{-1/2}\|\tilde X\|_F$ for traceless anti-Hermitian $\tilde{X}$, 
$\|\tilde C\|_F\le \|\tilde A\|_F+\|\tilde B\|_F$ holds. Moreover,
$|\tr C|=|\tr(A+B)|\le |\tr A|+|\tr B|$.
So, if we let $v(X):=(\|\tilde X\|_F,|{\rm tr} X|/\sqrt D)\in\mathbb R^2_{\ge 0}$, then 
$v_i(C)\le v_i(A)+v_i (B), \;i=\{1,2\}$ holds elementwise and thus
\[
\|C\|_F=\|v(C)\|_F \le \|v(A)+v(B)\|_F\le \|v(A)\|_F + \|v(B)\|_F=\|A\|_F+\|B\|_F,
\]
which proves Eq.~(\ref{eq:lemma1_cond3}).
\end{proof}

We now arrive at the following theorem by concatenating Eq. (\ref{eq:lemma1_cond1}) and letting the norm of each operator $A, B, ...$ be infinitesimal. 
\begin{theorem}\label{th:main_1}  
Given a unitary operator $U^\mathrm{(target)}$, which can be obtained as a solution of the Schr\"{o}dinger equation (\ref{eq:schrodinger_operator}) with initial condition (\ref{eq:initial_condition}) at some finite time $T\ge0$ with $H(t)$ in Eq. (\ref{eq:control_hamiltonian}), there exists a Hermitian operator $C[T]$ 
for any $T\geq 0$ such that
\begin{align}
\label{eq:main_3}
e^{-iC[T]} &= U^\mathrm{(target)} \;\mbox{and} \\
\label{eq:main_4}
i C[T]&\in \mathrm{Lie}\{iH_0,iH_1,\ldots,iH_M\},
\end{align}
where $C[T]$ satisfies
\begin{equation} \label{eq:main_5}
\|C[T]\|_F \leq \int_0^T dt \|H(t)\|_F. 
\end{equation}
\end{theorem}

Equation (\ref{eq:main_5}) in Theorem~\ref{th:main_1} allows us to have a phase-sensitive distance (metric) for a unitary group as follows: 
\begin{align}
\label{metric4unitaries}
d(U_1,U_2):=\min \left\{ \|C\|_F \left| e^{C}=U_1 U_2^{-1}\land {C\in\mathcal{L}}\right. \right\},
\end{align}
which is dependent on the algebra $\mathcal{L}$. 
A simple physical example, where this algebra-dependence can be seen, is given in Section IV of the Supplemental Material.

Before discussing applications of this metric, Eq. (\ref{metric4unitaries}), here is a brief remark on its geometric aspect.  Equation (\ref{eq:main_5}) also admits a geometric interpretation as an inequality for curve lengths on the unitary group: the right-hand side is the length of the control trajectory from $I$ to $U^\mathrm{(target)}$, whereas the LHS corresponds to the shortest such length. We return to this viewpoint in Sec. \ref{sec:conclusion}; our emphasis here is that Eq. (\ref{eq:main_5}) provides an explicit, operationally derived inequality that can be applied directly to obtain computable time lower bounds under realistic control constraints.

Lower bounds on gate complexity and control time have also been explored through inequalities that relate distances between unitary transformations to the Hamiltonians generating them~\cite{NielsenScience2006,NielsenPRA2006,Dowling2008,Arenz2017},\cite{Lee2018}. We now show how the metric $d(\cdot,\cdot)$ of Eq.~(\ref{metric4unitaries}) refines one such type of bound. These analyses are based on the inequality
\begin{align}\label{ineq_uni_neilsen}
    \left\|U_A(T)-U_B(T)\right\|\leq\int_{0}^T dt \left\|H_A(t)-H_B(t)\right\|
\end{align}
for the unitary operators $U_A(t)$ and $U_B(t)$ that are generated by Hamiltonians $H_A(t)$ and $H_B(t)$, respectively. This inequality holds with any unitarily invariant norm, including the Frobenius norm $\|\cdot \|_F$; thus, our result can be compared with it. 

It is possible to obtain a similar inequality in terms of the metric $d(\cdot,\cdot)$: 
\begin{align}\label{nielsenIneqWithMetric_d}
d(U_A(T),U_B(T)) &\leq \int_0^T dt \left\|H_A(t)-H_B(t)\right\|_F.
\end{align}
The LHS of Eq. (\ref{nielsenIneqWithMetric_d}) can be larger than that of Eq. (\ref{ineq_uni_neilsen}); that is, $d(\cdot,\cdot)$ may yield a larger lower bound for the RHS and thus the control time as well. The proof of Eq. (\ref{nielsenIneqWithMetric_d}), together with the unitary invariance of the metric $d(\cdot,\cdot)$, is given in Section V of the Supplemental Material, with careful treatment of the series convergence and its conditions for unitary operators. Since Eq. (\ref{ineq_uni_neilsen}) implies 
\begin{align}\nonumber
    \left\|e^{iC}-I\right\|_F \leq \left\|C \right\|_F,
\end{align}
we have
\begin{align}\label{compWithNielsenIneq}
    \left\|U_A(T)-U_B(T)\right\|_F =\left\|U_A(T) U_B(T)^{-1}-I \right\|_F
    \leq d(U_A(T) U_B(T)^{-1},I) = d(U_A(T), U_B(T)).
\end{align}
Since our theorem states that the metric $d(\cdot,\cdot)$ bounds the RHS of Eq. (\ref{ineq_uni_neilsen}) from below, there might be cases in which our bound is larger than those deduced from Eq. (\ref{ineq_uni_neilsen}). This can be seen quickly by considering $H_A(t)=\sigma_x$ and $H_B(t)=0$, where $\sigma_x$ is one of the standard Pauli matrices. Then, for $U_A(t)=\exp(-i\sigma_x t), U_B(t)=I$, the LHS of Eq. (\ref{compWithNielsenIneq}) is $2\sqrt{2}\sin(t/2)$, and the RHS is $\sqrt{2}t$, so the inequality holds for $t>0$.

Another interesting example of such bounds resulting from Eq. (\ref{ineq_uni_neilsen}) is that of Lee et al. \cite{Lee2018}:
\begin{equation}\label{eq_Lee_bound}
    T \ge \max_{V\in \bigcap_k\mathrm{Stab}(H_k )} \frac{\|[ U(T),V ]\|}{\|[ H_0,V ]\|}.
\end{equation}
Here, the total Hamiltonian $H(t)$ is assumed to have the same form as Eq. (\ref{eq:control_hamiltonian}).
Furthermore, in Eq. (\ref{eq_Lee_bound}), $\mathrm{Stab}(H_k)$ is a set of stabilizers for $H_k$, i.e., unitary operators that commute with $H_k$.
We fix the norm $\|\cdot\|$ as the Frobenius norm $\|\cdot \|_F$ to make it comparable with our bound.
We next show how Eq. (\ref{eq_Lee_bound}) can be refined. 
By combining Eqs. (\ref{nielsenIneqWithMetric_d}) and 
(\ref{compWithNielsenIneq}) and setting $H_A=H_0$ and $H_B=VH_0V^\dagger$ with arbitrary $V\in \bigcap _k \mathrm{Stab}(H_k)$, we obtain the following:
\begin{align}
    \| U(T) - V U(T)V^\dagger \|_F
    \le d\left (U(T), V U(T)V^\dagger \right ) & \le 
    \int_0^Tdt \ \| H (t) - VH(t) V^\dagger\|_F \nonumber \\
    &= T\cdot \| H_0 - VH_0 V^\dagger\|_F. \label{ineq4conjugatedHam}
\end{align}
Owing to the unitary invariance of the Frobenius norm, 
we have:
\begin{equation}
    T \ge  \frac{d\left (U(T), V U(T)V^\dagger \right )}{\| [H_0,  V]\|_F} \ge   \frac{ \| [U(T), V] \|_F}{\| [H_0,  V]\|_F}.
    \label{eq:refined_Lee}
\end{equation}
Since $V$ is an arbitrary unitary operation in $\bigcap _k \mathrm{Stab}(H_k)$, Eq. (\ref{eq:refined_Lee}) still naturally holds even if we take the maximum of the two expressions on the right side of Eq. (\ref{eq:refined_Lee}); thus, the bound obtained with our metric $d(\cdot,\cdot)$ (the first inequality of Eq. (\ref{eq:refined_Lee})) refines Lee et al.'s bound given in Eq. (\ref{eq_Lee_bound}) with the Frobenius norm.

As an application of Eq.~\eqref{eq:refined_Lee}, we consider a many-body control problem: a one-dimensional chain of $N$ spin-$1/2$ particles with XY interaction, following Ref.~\cite{Lee2018}.
Its drift Hamiltonian is given by
\begin{equation}
    H_0=J\sum _{j=1}^{N-1}\ket{j}\bra{j+1}+\ket{j+1}\bra{j},
\end{equation}
when we restrict our consideration to the single-excitation subspace. Here $\ket{j}\; (j=1, 2,..., N)$ represent the excited state at the $j$-th site. It is known that the system spanned by $\{ \ket{j} \}_{j=1}^N$ is fully controllable 
by accessing the Hamiltonian element $H_1=\ket{1}\bra{1}$ \cite{Schirmer2008,bmm2010}, and 
Lee et al. attempted to obtain the minimum necessary time to implement $U(T)=\exp (-i \pi/2 (\ket{1}\bra{N}+\ket{N}\bra{1}))$, a swap operator between the first and $N$-th spins \cite{Lee2018}.
The operator $V$ in Eq. (\ref{eq:refined_Lee}) is chosen to be $V=I-2\ket{1}\bra{1}$, which is clearly in  $\mathrm{Stab}(H_1)$. 
Then, by letting $H_0^\prime$ be a normalized Hamiltonian $H_0':=H_0/\|H_0\|_F=H_0/(J\sqrt{2(N-1)})$,  
\begin{equation}\label{eq_H0PrimeVH0PrimeVdagger}
\| H_0'- VH_0'V^\dagger \|_F =\frac{2J}{\|H_0 \|_F}\|(\ket{1}\bra{2}+\ket{2}\bra{1}) \|_F = \frac{2}{\sqrt{N-1}}
\end{equation}
and $VU(T)V^\dagger=U(T)^\dagger$ lead to
\begin{equation}\label{eq:ComparisonWLees}
\frac{ \| [U(T), V] \|_F}{\| [H_0',  V]\|_F}=
\frac{ \| U(T)-VU(T)V^\dagger \|_F}{\| H_0' - V H_0' V^\dagger\|_F}= 2\sqrt{2}\cdot \frac{\sqrt{N-1}}{2}=\sqrt{2(N-1)}.
\end{equation}
Meanwhile, $d(U(T),VU(T)V^\dagger)$ can be evaluated as
\begin{align}\label{eq example our new bound}
 d(U(T),VU(T)V^\dagger)& =
 d(U(T),U(T)^\dagger)\nonumber \\
 &=
 \min_{C\in \mathrm{su}(N)}
\left\{\|C\|_F \left|  e^C \right. =\exp(-i\pi(\ket{1}\bra{N}+\ket{N}\bra{1})\right\}\nonumber \\
&=\|-i\pi(\ket{1}\bra{N}+\ket{N}\bra{1}) \|_F
=\sqrt{2}\pi.
\end{align}
Then, Eq. (\ref{eq:refined_Lee}) is verified as 
\begin{equation}
    T \ge \frac{  d(U(T),VU(T)V^\dagger)}{\| [H_0',  V]\|_F}
    =\frac{\pi}{\sqrt{2}}\sqrt{N-1}
    >\sqrt{2(N-1)}=
    \frac{ \| [U(T), V] \|_F}{\| [H_0',  V]\|_F}.
\end{equation}
Thus, this example illustrates a realistic many-body setting in which the present bound gives a sharper estimate.

\section{Outline of proof}\label{sec:OutlineOfProof}
Let us delineate the proof of Theorem 
\ref{thm:main_dev_bound}, while its full details are given in Sections VI to IX of the Supplemental Material. Here, we first prove Lemma 
\ref{lem:bch_dev_triangle}
via the Baker-Campbell-Hausdorff (BCH) formula.
To this end, let us define an infinite series for operators $A$ and $B$ in a Lie algebra
\begin{align}\label{defM}
M(A,B):=A+B+\frac12[A,B]+\frac1{12}([A,[A,B]]+[B,[B,A]])+\cdots,
\end{align}
whose more precise form with higher-order commutators is shown in Section VI of the Supplemental Material. We also define the operator norm $\|\cdot\|_\mathrm{op}:Hom(\mathbb C^N)\rightarrow \mathbb R_{\geq 0}$ by
\begin{align}\label{operatorNorm}
\|A\|_\mathrm{op}:=\sqrt{\sup_{\ket x}\frac{\bra x A^\dagger A\ket x}{\langle x|x\rangle}}.
\end{align}
The BCH formula can be expressed as follows.
% Theorem 2 : BCH 
\begin{theorem}\label{th:BCH}
(Baker-Campbell-Hausdorff formula)
If $\|A\|_\mathrm{op}+\|B\|_\mathrm{op}<\log 2$, the series in $M(A,B)$ converges, and 
\begin{align}
e^{A}e^{B}&=e^{M(A,B)}\label{eq:BCH} \\
\mathrm{and}\;\; M(A,B)&\in\mathcal L(\{A,B\}) \label{eq:BCH_M}
\end{align}
hold.
\end{theorem}

When $A$ and $B$ are both anti-Hermitian operators, i.e., $A^\dagger = -A$, whose Frobenius norms are sufficiently small, we obtain a simple inequality that $\|M(A,B)\|_F$ should satisfy: 
\begin{equation}\label{eq:abs_rel_1}
\|M(A,B)\|_F^2\leq 2 \|A\|_F^2+2\|B\|_F^2-\|A-B\|_F^2.
\end{equation}
The proof of this inequality is somewhat lengthy and technical, so it is summarized in Section VII of Supplemental Material, along with the condition on the norm.

We are now ready to describe the proof of Lemma 
\ref{lem:bch_dev_triangle}
by using Theorem \ref{th:BCH} and Eq.~(\ref{eq:abs_rel_1}); the full proof is given in Section VIII of the Supplemental Material.
For anti-Hermitian operators $A$ and $B$, we choose large values of $m_a, m_b\in \mathbb{N}$ so that $(1/m_a)A$ and $(1/m_b)B$ are small enough for Eq. (\ref{eq:abs_rel_1}) to be used, and we let $m_a$ be odd. Let us define the operators $n=m_a+m_b$ by
\begin{equation}\label{def:Cjk}
\{C_j^{(1)}\} := \{ \overbrace{\frac{1}{m_a}A, \frac{1}{m_a}A, ..., \frac{1}{m_a}A,}^{m_a} \overbrace{\frac{1}{m_b}B, \frac{1}{m_b}B, ..., \frac{1}{m_b}B}^{m_b} \}
\end{equation}
for $j\in\{1,2,...,n\}$. For $k\in\mathbb{N}$, we recursively define a sequence of operators 
\renewcommand{\arraystretch}{2.2}
\begin{align}\label{def:Cjkplus1}
C_j^{(k+1)} &:=\left\{
\begin{array}{cc}
\displaystyle\frac{1}{2} M(C_{j-1}^{(k)},C_{j}^{(k)}) & \makebox{ if $j+k$ is odd and $j\neq 1$}, \\
\displaystyle\frac{1}{2} M(C_j^{(k)},C_{j+1}^{(k)}) & \makebox{ if $j+k$ is even and $j\neq n$}, \\
C_1^{(k)} & \makebox{ if $1+k$ is odd and $j= 1$}, \\
C_n^{(k)} & \makebox{ if $n+k$ is even and $j= n$}.
\end{array}
\right.
\end{align}
\renewcommand{\arraystretch}{1}

\begin{figure}
\includegraphics[width=75mm]{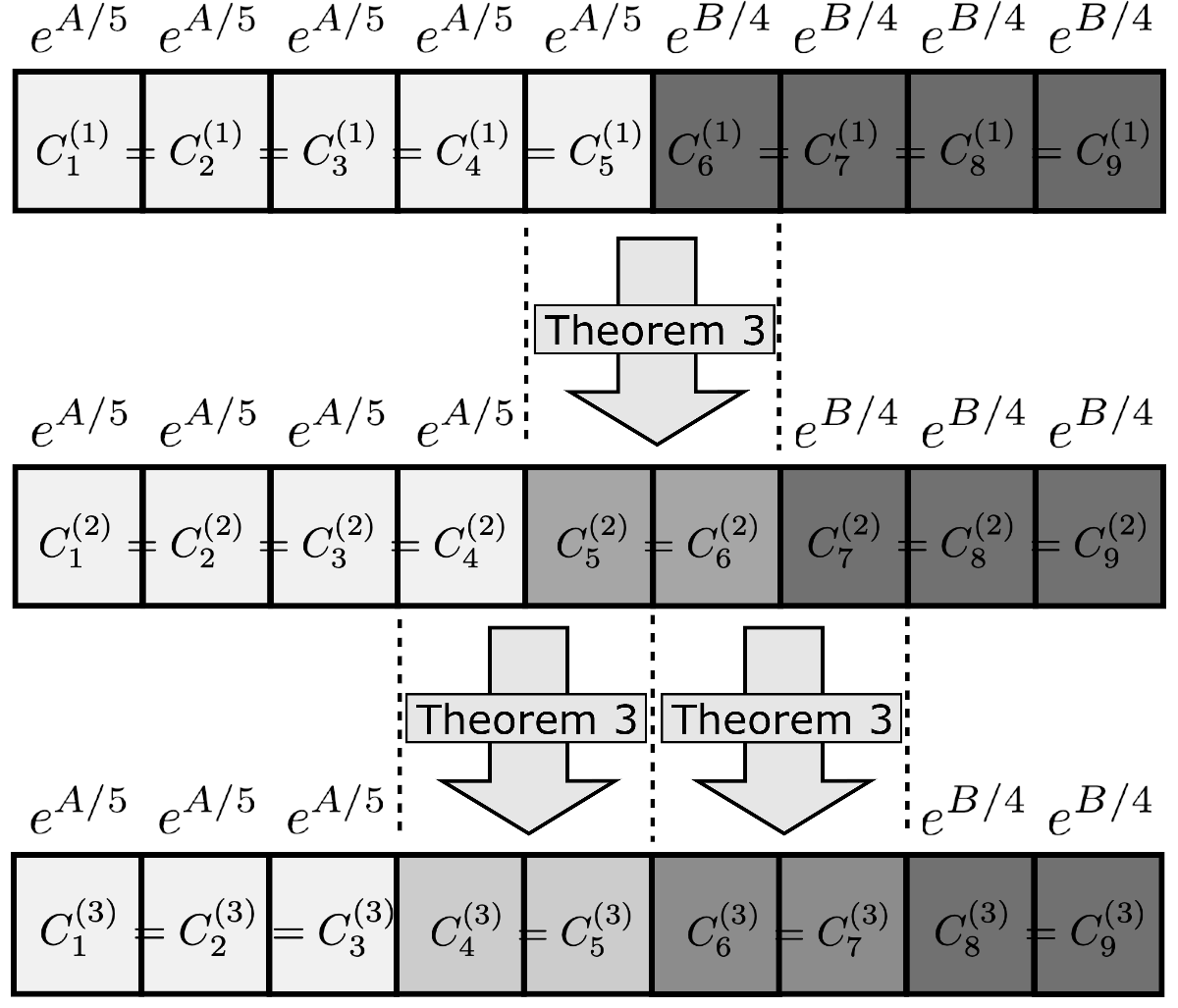}%2b from png thru photoshop, org by inkscape
\caption{Intuitive depiction of the evolution of the sequence $\{C_j^{(k)}\}$ as $k$ increases. Here, the overall operation is $e^A e^B$, and $e^A$ is divided into $5(=m_a)$ pieces of size $\exp(A/5)$, while $e^B$ is divided into $4(=m_b)$ pieces. Only the generator of the unitary operation, such as $C_1^{(1)}=A/5$, is shown in each box. Applying Theorem \ref{th:BCH} to the first row leads to the second row, synthesizing neighboring operators, namely, $C_5^{(2)}=C_6^{(2)}=M(A/5,B/4)/2$, while the others remain unchanged. Similarly, in the next step, we have 
$C_4^{(3)}=C_5^{(3)}=M(A/5,M(A/5,B/4)/2)/2$ and 
$C_6^{(3)}=C_7^{(3)}=M(M(A/5,B/4)/2,B/4)/2$. Repeated applications of Theorem \ref{th:BCH} make all the parts of the operations converge to a single unitary operation, i.e., $\exp(C/n)$, keeping the entire product equal to the original one: $(\exp(C/n))^n =e^C = e^A e^B$.}
\label{fig:def_Cjk}
\end{figure}

Note that most of the $C_j^{(k+1)}$ operators here are equal to $C_j^{(k)}$, particularly when $k$ is small, since in a sense, $C_j^{(k+1)}$ is an \textit{average} of the two neighboring operators in the $k$-th sequence. 
Figure 2 shows how two consecutive different unitary operators, e.g., $\exp(C_5^{(1)})$ and $\exp(C_6^{(1)})$ in the top row, may be averaged and how this averaging process propagates in the chain of $n$ operators. 

It can then be seen that, from Theorem \ref{th:BCH},
\begin{equation}\label{eq:eCeAeB}
\exp\left(C_1^{(k)}\right)
\exp\left(C_2^{(k)}\right)
\cdots \exp\left(C_{n}^{(k)}\right) = e^A e^B
\end{equation}
holds for all $k\in\mathbb{N}$. 
Furthermore, if we define two sequences
\begin{align}
\label{m_def:u_k}
\tilde u^{(k)}_1 &:= \sum_{j=1}^{n}\left(\dev  C_j^{(k)} \right)^2\\
\label{m_def:u_k_2}
\tilde u^{(k)}_2 &:= \sum_{j=1}^{n}\left|{\rm tr}  C_j^{(k)} \right|^2\\
\label{m_def:d_k}
\tilde d_1^{(k)} &:= \sum_{j=1}^{n-1} \left(\dev \left(C_{j}^{(k)}-C_{j+1}^{(k)}\right)\right)^2
\\
\label{m_def:d_k_2}
\tilde d_2^{(k)} &:= \sum_{j=1}^{n-1} \left|{\rm tr} \left(C_{j}^{(k)}-C_{j+1}^{(k)}\right)\right|^2
\end{align}
for $k>0$, then, as shown in Section VIII of the Supplemental Material, 
$\{\tilde{u}^{(k)}_1\}$ and $\{\tilde{u}^{(k)}_2\}$ is a nonincreasing sequence with respect to $k$, and 
$\tilde{d}^{(k)}_1$ and $\tilde{d}^{(k)}_2$ tend toward 0 as $k$ goes to infinity; i.e.,
\begin{align}
\label{m_eq:mon_u}
& \tilde u^{(1)}_1\geq \tilde u^{(2)}_1\geq \cdots \geq 0, \;\mathrm{and} 
\\
\label{m_eq:mon_u_2}
& \tilde u^{(1)}_2\geq \tilde u^{(2)}_2\geq \cdots \geq 0, \;\mathrm{and} 
\\
\label{m_eq:van_d}
& \lim_{k\rightarrow\infty} \tilde d^{(k)}_1 =0.
\\
\label{m_eq:van_d_2}
& \lim_{k\rightarrow\infty} \tilde d^{(k)}_2 =0.
\end{align}

Equations (\ref{m_eq:mon_u})--(\ref{m_eq:van_d_2}) imply that, for a sufficiently large $k$, $C_m^{(k)}$ contains a subsequence that converges to an operator, $(1/n)C$, regardless of $m$. Considering Eq. (\ref{eq:eCeAeB}), it is seen that the obtained operator $C$ is indeed the one that Lemma \ref{lem:bch_dev_triangle}  claims exists and satisfies $e^C=e^A e^B$. The full proof is given in Section VIII of the Supplemental Material.

The final step from Lemma \ref{lem:bch_dev_triangle} to Theorem \ref{thm:main_dev_bound} proceeds as follows.
(Theorem \ref{th:main_1} can also be proven from Lemma \ref{th:lemma_1} in a similar manner.)
For a given (Hermitian) Hamiltonian $H(t)$ for $0\le t\le T$, consider a sequence of Hamiltonians $\{H(\delta), H(2\delta),...,H(\lfloor(T/\delta)\rfloor \delta)\}$; roughly speaking, this sequence contains discretized $H(t)$ values with a short time interval $\delta$. Then, a Hermitian operator $C^{\delta}[T]$ that satisfies 
\begin{align}\label{eHs}
\exp\left(-iC^{\delta}[T]\right) &= \exp\left(-i H(\lfloor (T/\delta)\rfloor\delta)\delta\right) \cdots \exp\left(-i H(2\delta) \delta\right) \exp\left(-i H(\delta)\delta \right)
\end{align}
can be obtained by applying 
Lemma \ref{lem:bch_dev_triangle} (or Lemma \ref{th:lemma_1}) recursively. Letting $\delta\searrow 0$ makes $C^\delta[T]$ approach $C[T]$ in 
Theorem \ref{thm:main_dev_bound} (or Theorem \ref{th:main_1}), since the RHS of Eq. (\ref{eHs}) tends toward the $U_T^\mathrm{(target)}$ generated by $H(t)$. The proof is given in more detail in Section IX of the Supplemental Material.

\section{Discussion and conclusion}\label{sec:conclusion}
We have derived BCH-based inequalities for effective generators of controlled unitary evolutions. The main result gives a lower bound on the time required to implement a target unitary in terms of a projective operator distance and the time-integrated strength of the applied Hamiltonian. Since the deviation removes the trace component, the bound is insensitive to global phase and naturally defines a distance on the corresponding projective unitary group.

The bound incorporates the Lie-algebra membership constraint on the effective generator, but not the time overhead required to synthesize the minimizing logarithmic generator from the available Hamiltonians. It therefore rules out algebraically inadmissible shortcuts, while leaving open the finer problem of quantifying synthesis costs within the dynamical Lie algebra. We compared this projective operator-level bound with familiar QSL-type estimates and unitary-distance bounds, and found that it can give a sharper lower bound in the examples considered.

Let us make a remark on a geometric interpretation of the main results. The general terminology and notations in the following can be found in textbooks on geometry, e.g., \cite{gallotBook2004}. For a dynamical Lie algebra $\mathcal{L}$, we assume $e^{\mathcal{L}}$ is closed on the unitary group $U(D)$, where $D$ is a dimension of the quantum system. 
A Riemannian metric $g_I$ on identity $I$ can now be defined by a Hilbert-Schmidt inner product: $g_I(X,Y)=\mathrm{tr} X^\dagger Y$, where $X, Y \in \mathcal{L}$.
Then, it is known that we can extend $g_I$ to all the Lie group $e^{\mathcal{L}}$ so that 
the resulting Riemannian metric is bi-invariant. Here, Riemannian metric $g_U$ is called bi-invariant if for all $U,V\in e^\mathcal{L}$ and $X,Y \in T_Ue^\mathcal{L}$
\begin{equation}\label{eq bi invariance}
g_U(X,Y)=g_{VU}\big( (dL_V)_UX, (dL_V)_UY\big)=g_{VU}\big( (dR_V)_UX, (dR_V)_UY\big),
\end{equation}
where $T_U e^\mathcal{L}$ is a tangent space of $e^\mathcal{L}$ at $U$, $(dL_V)_U$ and $(dR_V)_U$ are differential maps of the left action $L_V$ and the right action $R_V$ of $e^\mathcal{L}$. Namely, $L_V(U)=VU, R_V(U)=UV$, and $(dL_V)_UX=VX, (dR_V)_UX=XV$ for $X \in T_Ue^\mathcal{L}$. 

Now, suppose that a curve $U(t)$ is a solution of Schr\"{o}dinger equation, Eq. (\ref{eq:schrodinger_operator}), connecting $U(0)=I$ and $U(T)=U^\mathrm{(target)}$. 
By writing a velocity vector of $U(t)$ as $X_{U(t)}:=dU/dt \in T_{U(t)}e^\mathcal{L}$, the pull-back of $X_{U(t)}$ to the identity is given by 
\begin{equation}\label{eq pull back}
X_I(t):=(dL_{U(t)^\dagger})_{U(t)}X_{U(t)}=U(t)^\dagger\frac{dU}{dt} = -iU(t)^\dagger H(t) U(t) \in \mathcal{L}.    
\end{equation}
Then, the length of the curve $\{U(t)\}_{t=0}^T$ is
\begin{align}\label{eq:length_gen_timeevolution}
\mbox{length}(\{U(t)\}_{t=0}^T) &=  \int _{0}^T \sqrt{g_U\big( X_{U(t)}, X_{U(t)}\big)} dt  
= \int _{0}^T \sqrt{g_I\big( X_I(t), X_I(t) \big)} dt \nonumber \\
&=\int _{0}^T \sqrt{\mathrm{tr} \big( X_I(t)^\dagger X_I(t) \big)} dt \nonumber \\
&=  \int _{0}^T \| H(t) \|_F dt.
\end{align}
%where Eqs. (\ref{eq bi invariance}) and (\ref{eq pull back}) are used in the second and fourth equalities.
This length must be lower-bounded by the length of the minimum geodesic connecting $I$ and $U^\mathrm{(target)}$.
An important observation is that under this bi-invariant metric $g_U$ of the compact Lie group $e^\mathcal{L}$, 
any geodesic through identity $I$ is given by 
an exponential map $\exp: t\mapsto e^{-itH}$, where $iH\in \mathcal{L}$.
In other words, for a given $U^{\mathrm{(target)}}$, there exists a time-independent Hamiltonian $iH_0\in \mathcal{L}$ such that the minimum geodesic between $I$ and $U^{\mathrm{(target)}}$ is given by $U_0(t)=e^{-itH_0}$. Then, if $U^{\mathrm{(target)}}$ is achieved at $t=T$, the length of the geodesic is evaluated by Eq. (\ref{eq:length_gen_timeevolution}) as 
\begin{align}\label{eq:length_min_geodesic}
\mbox{length}(\{U_0(t)\}_{t=0}^T) =  \int _{0}^T \| H_0 \|_F dt=T \|H_0\|_F,
\end{align}
and therefore, 
\begin{align}\label{eq:geo_main_ineq}
   T \|H_0\|_F \le  \int _{0}^T \| H(t) \|_F dt.
\end{align}
Identifying $C[T]=TH_0$,
we see Eq. (\ref{eq:geo_main_ineq}) is equivalent to Eq. (\ref{eq:main_5}).
Thus, Theorem \ref{th:main_1} can be understood as a consequence of the geometrical consideration of the space of unitaries under the assumption that $e^\mathcal{L}$ is closed in $U(D)$. 
However, this assumption may not hold, in general. We give an example, where 
$\mathcal{L}$ generates a non-compact group, in Section I of the Supplemental Material.

Despite the difficulties of analyzing quantum control time problems, their importance cannot be overemphasized, and our general arguments and results can be useful in exploring this rich and interesting field. A natural next step is to go beyond Lie-algebra membership and develop a more refined account of the cost of generating different directions within the dynamical Lie algebra. Such refinements may provide sharper lower bounds for genuinely many-body quantum control problems with limited access.

\section*{Author contributions}
All authors contributed to the development of the results, their interpretation, and the preparation of the manuscript. G.K. led the derivation and proof of the main theorem. M.O. and K.M. contributed to verification and refinement of the proof and to the interpretation of the results. K.M. coordinated and supervised the overall project. 

ChatGPT (OpenAI) was used solely for language editing, grammar checking, and improving readability. All scientific content was reviewed and approved by the authors, who take full responsibility for the manuscript.

\section*{Acknowledgments}
We acknowledge financial support from the JSPS, Japan Kakenhi (C) No. 20K03779 and No. 21K03388, and Japan Kakenhi (B) No. 23K25793. 
%MO also thanks for support by the JSPS, Japan Kakenhi (C) No. 16K00014.

% ===== Supplemental Material for arXiv =====
\clearpage

% Restart Supplement sections at I, II, ... rather than continuing from the main text.
\setcounter{section}{0}
\setcounter{subsection}{0}
\setcounter{subsubsection}{0}

% Restart supplementary equation numbering and use (S01), (S02), ...
\setcounter{equation}{0}
\renewcommand{\theequation}{S\ifnum\value{equation}<10 0\fi\arabic{equation}}

% Restart theorem-like counters as in the standalone Supplement.
\setcounter{lemma}{0}
\setcounter{definition}{0}
\setcounter{theorem}{0}
\setcounter{corollary}{0}
\setcounter{propostition}{0}

\begin{center}
\textbf{\large{SUPPLEMENTARY MATERIALS}}
\end{center}

\section{Examples outside the anti-Hermitian setting}

\label{app:remarksOnLemma1}
In Lemma \ref{lem:bch_dev_triangle}, it is essential that operators $A$ and $B$ be not only linear but also anti-Hermitian. Below is an example in which $e^{C}= e^{A}e^{B}$ never holds for $C\in \mathcal L(\{A,B\})$ with non-anti-Hermitian $A$ and $B$.

Let $A:= \frac{i\pi}{2}\left(
\begin{array}{cc}
1 & 1 \\ 0 & -1
\end{array}
\right)$ and $B:= \frac{i\pi}{2}\left(
\begin{array}{cc}
1 & -1 \\ 0 & -1
\end{array}
\right).$
Then, 
\begin{equation}\nonumber
\mathcal L(\{A,B\})=\left\{\left.
i\left(
\begin{array}{cc}
a&c\\
0&-a
\end{array}
\right)
\right|a\in\mathbb R,c\in\mathbb C\right\}
\end{equation}
and 
\begin{equation}\nonumber
e^{A}e^{B} = \left(
\begin{array}{cc}
-1 & 2 \\ 0 & -1
\end{array}
\right).
\end{equation}
The exponential form of an element of $\mathcal{L}(\{A,B\})$ is 
\begin{equation}\nonumber
\exp\left[{i\left(\begin{array}{cc}
a&c\\
0&-a
\end{array}
\right)}\right]
=
\left(\begin{array}{cc}
e^{ia}& \frac{ic}{a} \sin a\\
0&e^{-ia}
\end{array}
\right)
\end{equation}
for $a\neq 0$, and 
\begin{equation}\nonumber
\exp\left[i{
\left(\begin{array}{cc}
0&c\\
0&0
\end{array}
\right)}\right]
=
\left(\begin{array}{cc}
1&i c\\
0&1
\end{array}
\right).
\end{equation}
Therefore, $e^{C}$ cannot be equal to $e^{A}e^{B}$ for any $C\in \mathcal{L}(\{A,B\})$.

Furthermore, Eqs.~(\ref{eq:lemma_product}) and (\ref{eq:lemma_lie_membership}) state that $e^{A}e^{B}\in e^{\mathcal L(\{A,B\})}$ for anti-Hermitian operators $A$ and $B$; that is, $e^{\mathcal{L}}$ is a group. This statement is not necessarily obvious, although it may appear obvious at first sight.
If $e^{\mathcal L(\{A,B\})}$ is a compact set, it may be straightforward to deduce this assertion from a known fact, such as Theorem \ref{th:BCH}.
Nevertheless, the compactness of a Lie group cannot be guaranteed even if its generators are anti-Hermitian. 
For instance, 
\begin{align}\label{appA:noncompactG}
\exp\left[\mathcal L\left(\left\{\left(
\begin{array}{cc}
i&0\\
0&i\pi
\end{array}
\right) \right\}\right)\right]=\left\{
\left.
\left(
\begin{array}{cc}
e^{ia}&0\\
0&e^{ia\pi}
\end{array}
\right)
\right|a \in\mathbb R\right\}
\end{align}
forms a one-dimensional Lie group consisting of unitary matrices; however, it is not compact. To see the noncompactness of this group, imagine a matrix $\left(\begin{array}{cc} -1 & 0 \\ 0 & -1 \end{array}\right)$. It cannot be reached by the matrix in Eq. (\ref{appA:noncompactG}), but it is possible to approach it arbitrarily closely. This implies that if the group is not compact, the operator $M(A,B)$ in Eq. (\ref{defM}) may not converge in $\mathcal{L}(\{A,B\})$. This example shows that $\exp\mathcal L$ need not be compact, or equivalently need not be a closed Lie subgroup of $U(D)$. Consequently, one cannot rely on compactness alone to conclude that a limiting BCH generator, such as $M(A,B)$ in Eq.~\eqref{defM}, converges to an element of $\mathcal L$.
%%%%

\section{Example illustrating the possible discontinuity of $C[T]$}
\label{app:counter_example}
We give a specific example to show that $C[T]$ in Theorem \ref{th:main_1} cannot always be a continuous function of $T$. Suppose that a Hamiltonian $H(t)$ is given by
\begin{align}
H(t)=
\left\{
\begin{array}{cl}
\pi \cos t
\left(
\begin{array}{cc}
1& 0 \\ 0 & -1
\end{array}
\right)
&\makebox{ for $t<\frac12 \pi$ and $t\ge \pi$},
\vspace{2mm}\\
\pi \cos t
\left(
\begin{array}{cc}
0 & 1 \\ 1 & 0
\end{array}
\right)
&\makebox{ for $ \frac 12\pi\leq t <\pi$}.
\end{array}
\right.
\end{align}
The solution $U(t)$ of Eq. (\ref{eq:schrodinger_operator}) with the initial condition Eq. (\ref{eq:initial_condition}) is uniquely written as 
\begin{align}
U(t)=\left\{
\begin{array}{cl}
\exp\left(-i \pi \sin t
\left(
\begin{array}{cc}
1 & 0 \\ 0 & -1
\end{array}
\right)\right)
&\makebox{ for $t<\frac12 \pi$ and $t\ge \pi$},
\vspace{2mm}\\
\exp\left(-i \pi \sin t \left(
\begin{array}{cc}
0 & 1 \\ 1 & 0
\end{array}
\right)\right)
&\makebox{ for $ \frac 12\pi\leq t <\pi$}.
\end{array}\right.
\end{align}
Then, considering Eq. (\ref{eq:theorem_target_generator}), i.e., $e^{-iC[T]}=U^{(\mathrm{target})}$, $C[T]$ should satisfy
\begin{align}\label{ctapp}
C[T]=\left\{
\begin{array}{cl}
\left(
\begin{array}{cc}
\pi (\sin t +2n_1(t)) & 0 \vspace{2mm} \\ 0 & \pi (-\sin t + 2n_2(t))
\end{array}
\right)
&\makebox{ for $0<t<\frac12 \pi$ and $t\ge \pi$}, 
\vspace{3mm}\\
\left(
\begin{array}{cc}
\pi (n_3(t)+n_4(t)) & \pi (\sin t+n_3(t)-n_4(t)) \vspace{2mm} \\ \pi (\sin t + n_3(t)-n_4(t)) & \pi (n_3(t)+n_4(t))
\end{array}
\right)
&\makebox{ for $ \frac 12\pi< t<\pi$ },
\end{array}\right.
\end{align}
where $n_k(t)\in\mathbb N$. If $C[T]$ is continuous with respect to $t$, the nondiagonal elements, $n_3(t)$ and $n_4(t)$, of the second matrix should be constant, and they must approach zero when $t\searrow \pi/2$ and $t\nearrow \pi$. These requirements cannot be satisfied simultaneously for $n_k(t)\in \mathbb{N}$; thus, $C[T]$ is not necessarily continuous for all $t$.

%
% Appendix : Dependence of the metric (15) on algebra
%
\section{Dependence of the metric in Eq.~(\ref{eq:projective_distance}) on the algebra}
%(\ref{metric4unitaries})
\label{metricDepOnAlgebra}

We note that the metric defined by Eq. (\ref{eq:projective_distance}) for unitary operators depends on an algebra that has anti-Hermitian matrix representations. Specifically, while operators belonging to distinct algebras, $C_1\in L_1$ and $C_2\in L_2$, may generate the same unitary $U_1 U_2^{-1}$, i.e., $e^{C_1}=e^{C_2}=U_1 U_2^{-1}$, their minimum norms can differ. An example can be shown with the following two algebras, $L_1$ and $L_2$, and two unitaries, $U_1$ and $U_2$:
\begin{align}
L_1 &:=\mathcal L\left(\left\{
\left(
\begin{array}{ccc}
0 & i & 0 \\
i & 0 & 0 \\
0 & 0 & 0
\end{array}\right),
\left(
\begin{array}{ccc}
0 & 1 & 0 \\
-1 & 0 & 0 \\
0 & 0 & 0
\end{array}\right),
\left(
\begin{array}{ccc}
i & 0 & 0 \\
0 & -i & 0 \\
0 & 0 & 0
\end{array}\right),
\left(
\begin{array}{ccc}
i & 0 & 0 \\
0 & i & 0 \\
0 & 0 & -2i
\end{array}
\right)\right\}\right), \\
L_2 &:= \mathcal L\left(\left\{
\left(
\begin{array}{ccc}
i & 0 & 0 \\
0 & 2i & 0 \\
0 & 0 & -3i
\end{array}
\right)\right\}\right), \\
U_1 &:= \left(
\begin{array}{ccc}
-1 & 0 & 0 \\
0 & 1 & 0 \\
0 & 0 & -1
\end{array}
\right), \;\mbox{and}\;\;
U_2 :=I=
\left(
\begin{array}{ccc}
1 & 0 & 0 \\
0 & 1 & 0 \\
0 & 0 & 1 
\end{array}\right).
\end{align}
Then, we have the operator $C$ for each algebra
\begin{align}
\left\{ C_1\left| e^{C_1}=U_1 U_2^{-1}\;\land\;C_1\in L_1 \right. \right\} &=
\left\{ \left( \left. 
\begin{array}{ccc}
(2n+1)i\pi & 0 & 0 \\
0 & 2mi\pi & 0 \\
0 & 0 & -(2n+2m+1)i\pi
\end{array}
\right) \right| n,m\in\mathbb Z \right\}, \\
\left\{ C_2 \left| e^{C_2}=U_1 U_2^{-1}\;\land\;C_2 \in L_2 \right. \right\} &=
\left\{ \left( \left. 
\begin{array}{ccc}
(2n+1)i\pi & 0 & 0 \\
0 & 2(2n+1)i\pi & 0 \\
0 & 0 & -3(2n+1)i\pi
\end{array}
\right) \right| n\in\mathbb Z \right\}.
\end{align}
Thus, the metrics $\tilde d_1(U_1, U_2)$ and $\tilde d_2(U_1, U_2)$ under algebras $L_1$ and $L_2$ are 
\begin{align}
\tilde d_1(U_1, U_2) &= \min_{L_1}\{\dev C_1\} = \dev \left(
\begin{array}{ccc}
\pi & 0 & 0 \\
0 & 0 & 0 \\
0 & 0 & -\pi
\end{array} \right) = \sqrt{2}\pi, \; \mbox{and } \\
\tilde d_2(U_1, U_2) &= \min_{L_2}\{\dev C_2\} = \dev \left(
\begin{array}{ccc}
\pi & 0 & 0 \\
0 & 2\pi & 0 \\
0 & 0 & -3\pi
\end{array}\right) = \sqrt{14}\pi,
\end{align} 
respectively, and are thus different, as noted.

\section{A physical example that leads to algebra-dependence of the distance Eq. (\ref{metric4unitaries})}
Suppose two systems, each of which is comprised of 9 spins-1/2 and is governed by the following Hamiltonians.
\begin{align}
H_1&=\sum_{j=1}^{3}\sum_{j'=j+1}^{4}g_1\vec\sigma_j\vec\sigma_{j'}
    \\
H_2&=\sum_{j=1}^{3}\sum_{j'=j+1}^{4}g_2\vec\sigma_j\vec\sigma_{j'}+\sum_{j=5}^{8}\sum_{j'=j+1}^{9}g_2\vec\sigma_j\vec\sigma_{j'},
\end{align}
where $\vec\sigma_j:=(\sigma_j^{(x)},\sigma_j^{(y)},\sigma_j^{(z)})$ is a standard set of Pauli operators, with $\{\ket{\uparrow},\ket{\downarrow}\}$ as eigenstates, acting on the $j$-th spin. The Hamiltonian $H_1$ describes the dynamics of a set of four spins, whose interaction graph is fully connected, and the remaining five spins are not interacting with others. $H_2$ is for two sets of spins; one is the same four-spin set as in $H_1$ (with a different coupling strength) and the other is for a similar fully-connected five-spin graph. 

Let $\ket{j}$ denote a state, where only the $j$-th spin is in $\ket{\uparrow}$, and let us consider a state space $\mathcal{H}$ spanned by 
$\{4^{-1/2}\sum_{j=1}^{4}\ket{j},5^{-1/2}\sum_{j=5}^{9}\ket{j}\}$, namely so-called first-excitation subspace. The above Hamiltonians have matrix representations in $\mathcal{H}$ as
\begin{equation*}
   \bar{H}_1 =g_1 \left(
   \begin{array}{cc}
       3 & 0  \\
        0& 0 
   \end{array}
   \right), \;\;
   \bar{H}_2 =g_2\left(
   \begin{array}{cc}
       3 & 0  \\
        0& 4 
   \end{array}
   \right).
\end{equation*}

Now let us compute the metric defined by Eq. (\ref{metric4unitaries}) between  
\begin{equation*}
U_1= \left(
\begin{array}{cc}
-1 & 0 \\
0 & 1 
\end{array}
\right), \;\mbox{and}\;\;
U_2 =I=
\left(
\begin{array}{cc}
1 & 0  \\
0 & 1  
\end{array}\right),
\end{equation*}
with corresponding algebras $L_1 =\mathcal L(\{ i\bar H_1\})$ and $L_2= \mathcal L(\{i\bar H_2\})$. 

Then, we have the operator $C$ for each algebra 
\begin{align}
\left\{ C_1\left| e^{C_1}=U_1 U_2^{-1}\;\land\;C_1\in L_1 \right. \right\} &=
\left\{ \left. \left( 
\begin{array}{cc}
(2n+1)i\pi & 0 \\
0 & 0 
\end{array}
\right)
\right| n\in\mathbb Z \right\}, \\
\left\{ C_2 \left| e^{C_2}=U_1 U_2^{-1}\;\land\;C_2 \in L_2 \right. \right\} &=
\left\{ \left. \left(  
\begin{array}{cc}
3(2n+1)i\pi & 0 \\
0 & 4(2n+1)i\pi  
\end{array}
\right)
\right| n\in\mathbb Z \right\}.
\end{align}
The metrics $d_1(U_1, U_2)$ and $d_2(U_1, U_2)$ under algebras $L_1$ and $L_2$ are 
\begin{align}
d_1(U_1, U_2) &= \min_{L_1}\{\|C_1\|_F\} = \left\| \left(
\begin{array}{cc}
\pi & 0 \\
0 & 0
\end{array} \right)\right\|_F = \pi, \; \mbox{and } \\
d_2(U_1, U_2) &= \min_{L_2}\{\|C_2\|_F\} = \left\| \left(
\begin{array}{cc}
3\pi & 0 \\
0 & 4\pi 
\end{array}\right)\right\|_F = 5\pi,
\end{align} 
respectively, and are thus different, as noted. The lower bounds, $T_1, T_2$, of time to realize $U_1$ are
\begin{align}
T_1&=\frac{d_1(U_1,U_2)}{\left\|H_1\right\|_F}=\frac{\pi}{3g_1},
\\
T_2&=\frac{d_2(U_1,U_2)}{\left\|H_2\right\|_F}=\frac{\pi}{g_2},
\end{align} 
which implies that the lower bound is dependent on the algebras. Also, just by exponentiating the Hamiltonians for these times, it is straightforward to see that these bounds are indeed achievable.

%
% Appendix : Proof of Eq. (\ref{nielsenIneqWithMetric_d}) 
%
\section{Proof of Eq. (\ref{nielsenIneqWithMetric_d})}
\label{app:nielsenIneq}
Since $d(\cdot,\cdot)$ fulfills the requirements for a metric, it would be possible to prove Eq. (\ref{nielsenIneqWithMetric_d}) by using the triangle inequality, as was done for Eq. (\ref{ineq_uni_neilsen}) in \cite{NielsenScience2006,NielsenPRA2006}. Nevertheless, we aim to prove inequality (\ref{nielsenIneqWithMetric_d}) here by carefully considering the uniformity of convergence in terms of time rather than relying on intuitions about the triangle inequality. 

\setcounter{lemma}{2} 
\begin{lemma}\label{UnitaryInvarianceOfd}
The metric $d(U_1,U_2)$ is invariant under unitary operations $V\in e^{\mathcal L}$; i.e.,
\begin{align}
d(U_1, U_2) = d(V_1U_1V_2,V_1 U_2V_2)
\end{align}
for any $U_1,U_2,V_1,V_2\in e^{\mathcal L}$.
\end{lemma}

\begin{proof}
Since $V_1^{-1}CV_1\in {\mathcal L}$ holds for any $C\in\mathcal L$, we have $\|C\|_F=\|V_1^{-1}C V_1\|_F$ and 
\begin{align}
d(U_1, U_2) &= \min\left\{\|C\|_F \left| e^{C}=U_1 U_2^{-1}\right.\right\} \nonumber \\
&= \min \left\{\|V_1^{-1}C V_1\|_F \left| e^{V_1^{-1}C V_1}=U_1 U_2^{-1}\right.\right\} \nonumber \\
&= \min \left\{\|C\|_F \left| e^{C}=V_1 U_1 U_2^{-1} V_1^{-1} \right.\right\} \nonumber \\
&= \min \left\{\|C\|_F \left| e^{C}=(V_1 U_1 V_2)(V_1 U_1 V_2)^{-1} \right.\right\} \nonumber \\
&= d(V_1U_1V_2,V_1 U_2V_2). \nonumber 
\end{align}
\end{proof}

\begin{lemma}
\label{lem:small_distance}
Suppose that $H(t)$ is a bounded Hermitian operator that acts on a $D$-dimensional space and is a piecewise continuous function of $t$.
A unitary operator $U(t)$ is a solution of the differential equation Eq.~(\ref{eq:schrodinger_operator}) under the initial condition Eq.~(\ref{eq:initial_condition}).
Then, there is a $T$-parameterized Hermitian operator $C[T]$ such that Eqs. (\ref{eq:main_3}) and (\ref{eq:main_4}), as well as
\begin{align}
\left\|C[T]+\int_0^T dt H(t)\right\|_{F} &\leq 4\sqrt D\alpha^2 T^2, \; \mbox{and}
\label{tmp:lem_eq_1}\\
d(U(T),I) &= \left\|C[T]\right\|_F, 
\label{tmp:lem_eq_2}
\end{align}
are satisfied for $T$. 
%Editor: Please ensure that the intended meaning has been maintained in the following edit.
The range of this operator 
is constrained by 
\begin{align}
\alpha T &\leq \frac 13,
\label{ass:small_distance_10} \\
\beta T &\leq \pi.
\label{ass:small_distance_2}
\end{align}
with 
\begin{align}\label{ass:small_distance_1}
\alpha = \max_{0\leq t\leq T}\left\| H(t)\right\|_\mathrm{op},\; \mathrm{and}\;\;
\beta = \max_{0\leq t\leq T}\left\| H(t)\right\|_{F}.
\end{align}
\end{lemma}
The factor $1/3$ in Eq. (\ref{ass:small_distance_10}) is a small number chosen arbitrarily to ensure that the series in the following argument will be convergent. The definition of the operator norm $\|\cdot \|_\mathrm{op}$ is 
\begin{align*}
\left\|A \right\|_\mathrm{op}=\sqrt{\sup_{\ket x}\frac{\bra x A^\dagger A\ket x}{\langle x|x\rangle}},
\end{align*}
which is also given in Eq. (\ref{operatorNorm}).

\begin{proof}
The formal expansion (Dyson series) of $U(T)$ is 
\begin{align}
U(T) &= I-i\int_0^T dt H(t) - \int_0^T dt H(t)\int_0^{t} dt' H(t')
+i\int_0^T dt H(t) \int_0^{t} dt' H(t') \int_0^{t'} dt'' H(t'') +\cdots.
\label{eq:formal_integral}
\end{align}
Note that the RHS of Eq. (\ref{eq:formal_integral}) converges for an arbitrary $T$. This is because $||H(T)||_\mathrm{op}$ is bounded from above by $\alpha$ for $0\le t\le T$; hence, the operator norm of the $n$-th term is smaller than $(1/n!)(\alpha t)^n$. 
Let $W(t)$ be the RHS of Eq. (\ref{eq:formal_integral}) except for the first two terms, i.e.,
\begin{align}
W(T) &:= U(T)-I+i\int_0^T dt H(t).
\label{def:W_t}
\end{align}
Then, since the operator norm of the $n$-th term of Eq. (\ref{eq:formal_integral}) is bounded by $(\alpha T)^n$, we have 
\begin{align}
\|W(T)\|_\mathrm{op} &\leq \frac{\alpha^2 T^2}{1-\alpha T}.
\label{eq:upper_bound_W_t}
\end{align}

Equations (\ref{ass:small_distance_10}) and (\ref{eq:upper_bound_W_t}) imply that $||i\int_0^T dt H(t)-W(T)||_{\rm op}\leq 1/2$; thus,
\begin{equation}\label{itildec}
i\tilde{C}(T):=-\sum_{n=1}^\infty\frac{1}{n}\left(i\int_0^T dt H(t)-W(T)\right)^n
\end{equation}
converges. Its RHS is equal to
\begin{equation}
\log\left(I-i\int_0^T dt H(t) + W(T)\right) = \log U(T);
\end{equation}
hence,
\begin{align}
U(T)&=\exp(i\tilde C(T)).
\end{align}
Furthermore, we have
\begin{align}
\left\|\tilde C(T)+\int_0^T dt H(t)\right\|_\mathrm{op} 
&=
\left\|W(T)+\sum_{n=2}^\infty\frac in(i\int_0^T dt H(t)-W(T))^n\right\|_\mathrm{op} 
\nonumber\\
&\leq
\left\|W(T)\right\|_\mathrm{op}+\sum_{n=2}^\infty\frac 1n
\left(\left\|i\int_0^T dt H(t)\right\|_\mathrm{op}+\left\|W(T)\right\|_\mathrm{op}\right)^n 
\nonumber\\
&\leq
\frac{\alpha^2 T^2}{1-\alpha T}+\sum_{n=2}^\infty\frac 1n
\left(\alpha T+\frac{\alpha^2 T^2}{1-\alpha T}\right)^n 
\nonumber\\
&= -\log\left(1-\frac{\alpha T}{1-\alpha T}\right)
-\alpha T \nonumber \\
&\leq 4\alpha^2 T^2,
\label{eq:small_distance_1} 
\end{align}
where we substitute Eq. (\ref{itildec}) to obtain the first equality and use the triangle inequality and $\left\|A^2\right\|\leq\left\|A\right\|^2$ to obtain the second line. The last inequality can be derived by noting that $-\log(1-x/(1-x))-x\leq 4 x^2$ holds for $0\le x< 0.383\cdots$, which is larger than the range of our assumption, Eq. (\ref{ass:small_distance_10}).

An inequality for $\|\tilde{C}(t)\|_\mathrm{op}$ can be obtained as 
\begin{align}
\left\|\tilde C(T)\right\|_\mathrm{op} 
&\leq
\left\|\tilde C(T)+\int_0^T dt H(t)\right\|_\mathrm{op} 
+\left\|\int_0^T dt H(t)\right\|_\mathrm{op} 
\nonumber\\
&\leq -\log\left(1-\frac{\alpha T}{1-\alpha T}\right)
\nonumber\\
&< \frac{3\pi}{4}.
\label{eq:small_distance_2}
\end{align}
The first line is a triangle inequality, and the second line is from Eq. (\ref{eq:small_distance_1}) under the assumption of Eq. (\ref{ass:small_distance_10}). The last inequality can be verified by noting that $-\log(1-x/(1-x))< 3\pi/4$ for $0\le x \le 1/3$.

With Eq. (\ref{eq:small_distance_2}), we can verify the following: 
\begin{align}
d(U(T),I) &= \min 
\left\{ \|C\|_F\left| e^{iC}=e^{i\tilde C}\;\land \; C\in\mathcal L \right.\right\},
\label{eq:small_distance_3}
\\
\sqrt{\|\tilde C\|_F^2+(2\pi - \|\tilde C\|_\mathrm{op})^2 - \|\tilde C\|_\mathrm{op}^2} &\leq \min 
\left\{ \|C\|_F \left| e^{iC}=e^{i\tilde C}\;\land \; C\in\mathcal L\;\land \;
C\neq \tilde{C} \right.\right\}=:d_1(T),
\label{eq:small_distance_4}
\end{align}
The LHS of Eq. (\ref{eq:small_distance_4}) is the Frobenius norm of $\tilde{C}$ after its largest eigenvalue, $\|C\|_\mathrm{op}$, is shifted by $2\pi$ so that it still generates the same $U(t)$. The equality in Eq. (\ref{eq:small_distance_4}) holds when $\mathcal{L}=\mathfrak{su}(N)$; otherwise, it does not necessarily hold.

Combining these considerations,
\begin{align}\label{dIsSmallerThand1}
d(U(T),I) &\leq \int_0^T dt \left\|H(t)\right\|_{F}\leq \pi
< \sqrt{(2\pi - \|\tilde C\|_\mathrm{op})^2 - \|\tilde C\|_{op}^2}\leq d_1(T)
\end{align}
is obtained.
The first inequality is due to the definition of the distance $d(\cdot,\cdot)$ and Theorem \ref{th:main_1}.
The second inequality comes from the condition 
Eq. (\ref{ass:small_distance_2}).
The third inequality holds because of Eq. (\ref{eq:small_distance_2}), i.e., $\|\tilde C[T]\|_{op}<3\pi/4$, and the last inequality is due to Eq. (\ref{eq:small_distance_4}).

To show Eq.~(\ref{tmp:lem_eq_2}), consider $C^\prime[T]$ in the dynamical Lie algebra $\mathcal{L}$, which satisfies $\exp(i C^\prime[T])=U(T)=\exp(i \tilde{C}[T])$ and $d(U(T),I)=\|C^\prime [T]\|_F < d_1$,
where the inequality is obtained from Eq. (\ref{dIsSmallerThand1}). From Eq.~(\ref{dIsSmallerThand1}), the RHS of Eq.~(\ref{eq:small_distance_3}) is strictly smaller than the RHS of Eq.~(\ref{eq:small_distance_4}). Thus, there are strictly fewer elements of the set in Eq.~(\ref{eq:small_distance_4}) than in Eq.~(\ref{eq:small_distance_3}), and the element that is included in the set in Eq.~(\ref{eq:small_distance_3}) but not in Eq.~(\ref{eq:small_distance_4}) is $\tilde{C}$. Since $\|C^\prime[T]\|_F<d_1$ implies that $C^\prime[T]$ is not contained in the RHS of Eq.~(\ref{eq:small_distance_4}), $\tilde{C}[T] = C^\prime [T]$. Hence, $d(U(T),I)=\|C[T]\|_F$, which is Eq.~(\ref{tmp:lem_eq_2}). 

Additionally, recalling the trivial relation
\begin{align}
\|A\|_{F}\leq \sqrt D\|A\|_\mathrm{op},
\label{eq:norm_relation}
\end{align}
where $D$ is the dimension of the space on which $A$ acts, Eq.~(\ref{eq:small_distance_1}) implies Eq.~(\ref{tmp:lem_eq_1}). 
\end{proof}

\begin{corollary}
\label{col:nielsenIneqWithD}
Suppose that $H_A(t)$ and $H_B(t)$ are $t$-parameterized bounded Hermitian operators and that they are piecewise continuous with respect to $t$. 
Additionally, $U_A(t)$ and $U_B(t)$ are unitary operators that follow the differential equation (\ref{eq:schrodinger_operator})
under the initial condition (\ref{eq:initial_condition}).
Then, Eq. (\ref{nielsenIneqWithMetric_d}) in the main text, namely,
\begin{align}
d(U_A(T),U_B(T)) &\leq \int_0^T dt \left\|H_A(t)-H_B(t)\right\|_F,
\label{eq:main_co_5}
\end{align}
holds for all $T\geq 0$.
\end{corollary}
\begin{proof}
Let us define
\begin{align}
\alpha&:=
\max_{0\leq t\leq T}\;
\max\left(\left\|H_A(t)\right\|_\mathrm{op},\left\|H_B(t)\right\|_\mathrm{op}\right),
\\
\beta&:=
\max_{0\leq t\leq T}\;
\max\left(\left\|H_A(t)\right\|_{F},\left\|H_B(t)\right\|_{F}\right),
\end{align}
as well as an integer $N$ such that 
\begin{align}
6\alpha T &< N,
\label{eq:main_co_2} \; \mbox{and}
\\
\frac 2\pi \beta T &< N.
\label{eq:main_co_1}
\end{align}

We now consider the Schr\"{o}dinger equation $idU(t)/dt=H(t)U(t)$ under the initial condition $U(0)=I$.
The Hamiltonian is given as follows for an interval $t \in [0,2\Delta_N)$ with $\Delta_N=T/N$ and $n\in\{1,...,N\}$:
\begin{align}\label{ham_pieces}
\tilde H_n(t):=
\left\{
\begin{array}{ll}
\displaystyle H_A \left((n-1)\Delta_N+t \right) & \; \mbox{if} \;\; \displaystyle 0\le t <\Delta_N\\
\displaystyle -H_B \left((n+1)\Delta_N - t \right) & \; \mbox{if} \;\; \displaystyle \Delta_N\le t < 2\Delta_N .
\end{array}
\right.
\end{align}
This Hamiltonian $\tilde{H}_n(t)$ evolves the system forward by $H_A(t)$ for a duration $\Delta_N$ and then backward by $H_B(t)$ for the same $\Delta_N$. Thus, the overall evolution over the duration $2\Delta_N$ shows the difference between these two Hamiltonians, and by accumulating such differences over $T$, we intend to prove Eq. (\ref{eq:main_co_5}).

The solution of the Schr\"{o}dinger equation with the Hamiltonian in Eq. (\ref{ham_pieces}) is
\begin{align}
    U^{(n)}(t)=
\left\{
\begin{array}{ll}
\displaystyle  U_A\left((n-1)\Delta_N + t \right) U_A^{-1}\left((n-1)\Delta_N \right)
&\makebox{if} \;\;  0\le t < \Delta_N \\
\displaystyle  U_B\left((n+1)\Delta_N - t \right) U_B^{-1}\left(n\Delta_N \right) U_A\left(n\Delta_N \right) U_A^{-1}\left((n-1)\Delta_N \right)
&\makebox{if} \;\; \Delta_N \le t < 2\Delta_N,
\end{array}
\right.
\end{align}
so at $t=0$, i.e., at the beginning of each small time interval $\Delta_N$, $U(0)=I$.
Therefore, at time $2\Delta_N=2T/N$, 
\[
U^{(n)}\left(2\Delta_N\right) = U_B\left((n-1)\Delta_N\right) U_B^{-1}\left(n\Delta_N \right) U_A\left(n \Delta_N\right) U_A^{-1}\left((n-1)\Delta_N\right),
\]
and the assumptions of Lemma \ref{lem:small_distance} are satisfied when 
the time $t$ in Eqs. (\ref{ass:small_distance_10}) and (\ref{ass:small_distance_2}) is set to $2\Delta_N$.
Therefore, Eq. (\ref{tmp:lem_eq_2}) in Lemma \ref{lem:small_distance} now implies that a $C[\cdot]$ exists such that
\begin{equation}
d\left(U_B\left((n-1)\Delta_N \right) U_B^{-1}\left(n \Delta_N\right) U_A\left(n \Delta_N\right) U_A^{-1}\left((n-1)\Delta_N \right), I\right) =\left\| C\left[2\Delta_N\right]\right\|_F,
\end{equation}
and
\begin{align} 
\left\| C\left[2\Delta_N\right]\right\|_F
&\leq \left\| \int_0^{2\Delta_N} dt \tilde H_n(t) \right\|_F
+
\left\| C\left[2\Delta_N \right] + \int_0^{2\Delta_N} dt \tilde H_n(t) \right\|_F
\nonumber\\
&\leq \left\| \int_0^{2\Delta_N} dt \tilde H_n(t) \right\|_F
+ 4\sqrt D \left(2\alpha \Delta_N\right)^2,
\label{eq:main_co_4}
\end{align}
using the triangle inequality and Eq. (\ref{tmp:lem_eq_1}). 
The first term in the last line of Eq. (\ref{eq:main_co_4}) can be upper bounded:
\begin{align}
\left\| \int_0^{2\Delta_N} dt \tilde H_n(t) \right\|_F
&=
\left\| \int_{(n-1)\Delta_N}^{n\Delta_N} dt 
\left( H_B(t) - H_A(t) \right) \right\|_F
\nonumber\\
&\leq
\int_{(n-1)\Delta_N}^{n\Delta_N} dt
\left\| H_B(t) - H_A(t) \right\|_F.
\label{eq:proof_col_3}
\end{align}

Combining these relations, we can evaluate the distance as follows:
\begin{align}\label{eq:proof_col_3ub}
d(U_A(t),U_B(t)) &= d(U_B^{-1}(t)U_A(t), I) \nonumber\\
&\leq \sum_{n=1}^{N}d\left( U_B^{-1}\left(n\Delta_N \right) U_A\left(n\Delta_N \right), \;U_B^{-1}\left((n-1)\Delta_N\right) U_A\left((n-1)\Delta_N\right) \right)
\nonumber\\
&= 
\sum_{n=1}^{N}
d\left(U_B\left((n-1)\Delta_N \right) U_B^{-1}\left(n\Delta_N \right) U_A\left(n\Delta_N \right) U_A^{-1}\left((n-1)\Delta_N\right), I \right)
\nonumber\\
&\leq
\sum_{n=1}^{N} \left( \left\|\int_0^{2\Delta_N} dt
\tilde H_n(t)\right\|_F + 4\sqrt D\left(2\alpha\Delta_N \right)^2 \right)
\nonumber\\
&\leq 
\int_{0}^{T}
dt \left\| H_B(t) - H_A(t) \right\|_F + 16\sqrt D \frac{\alpha^2 T^2}{N}.
\end{align}
The first and third lines are due to the unitary invariance of the distance $d(\cdot,\cdot)$ shown in Lemma \ref{UnitaryInvarianceOfd}. The second line 
%Editor: Please ensure that the intended meaning has been maintained in the following edit.
again comes from 
recursive applications of the triangle inequality. The last two inequalities are a result of Eqs. 
 (\ref{eq:main_co_4}) and (\ref{eq:proof_col_3}), respectively.
By letting $N$ tend toward infinity, the second term in the last line of Eq. (\ref{eq:proof_col_3ub}) vanishes; hence, Eq. (\ref{eq:main_co_5}) holds.

\end{proof}

%
% Appendix : Definition of M
%
\section{Definition of $M(\cdot ,\cdot)$}
\label{app:def_M}
The coefficients in the infinite series for $M(\cdot ,\cdot)$ are defined in the following manner.
\begin{definition}
For $\vec{c}=(c_1,c_2,...,c_n)\in \{0,1\}^{\otimes n}$, the functions $f(\vec c),g(\vec c),h(\vec c):\{0,1\}^{\otimes n}\rightarrow \mathbb R$ are those that satisfy the following equations:
\begin{align}
\sum_{n=1}^{\infty}\frac{(-1)^{n-1}}{n}
\sum_{n=0}^\infty\sum_{\vec c\in\{0,1\}^{\otimes n}}f(\vec c)A^{1-c_1}B^{c_1}A^{1-c_2}B^{c_2}\cdots A^{1-c_n}B^{c_n}
&:= \left(\left(\sum_{j=0}^\infty\frac{A^j}{j!}\right)\left(\sum_{j=0}^\infty\frac{B^j}{j!}\right)-I\right)^{n-1},
\label{def:f}
\end{align}
\begin{align}
(n+2)g(\vec c) &:=(-1)f(c_1,c_2,\cdots,c_n,1)-(-1)^n f(1-c_1,1-c_2,\cdots,1-c_{n},1),
\label{def:g}
\\
h(\vec c) &:=
2g(1,c_1,c_2,\cdots ,c_n)
-2g(0,c_1,c_2,\cdots ,c_n)
\nonumber\\
&\quad +\sum_{m=0}^n
(-1)^m g(c_m,c_{m-1},\cdots ,c_1)
g(c_{m+1},c_{m+2},\cdots ,c_n).
\label{def:h}
\end{align}
where $A$ and $B$ are noncommutative algebraic elements.
\end{definition}
Additionally, for a vector $\vec{c}$, we define $|\vec c|$ as
\begin{align}
|\vec c| &:=\sum_{m=1}^n c_m,
\end{align}
and we assume that $\sum_{\vec c\in \{0,1\}^{\otimes 0}}g(\vec c) = g(\varnothing)$, where $\varnothing$ is a null string.

Some of the specific values of these functions are: 
\begin{align}
\begin{array}{clcclcclccl}
f(\varnothing) = 1,
&\quad&
f(0) =-\displaystyle \frac{1}{2},
&\quad&
f(1) =-\displaystyle  \frac12, &\quad& 
\vspace{2mm}\\
f(0,0) =\displaystyle \frac1{12},
&\quad&
f(0,1) =-\displaystyle \frac1{6},
&\quad&
f(1,0) =\displaystyle \frac1{3},
&\quad&
f(1,1) =\displaystyle \frac1{12},
\vspace{2mm}\\
g(\varnothing) =\displaystyle \frac12,
&\quad&
g(0) =\displaystyle \frac1{12},
&\quad&
g(1) =-\displaystyle \frac1{12},
&\quad&
h(\varnothing) =-\displaystyle \frac1{12},...
\end{array}
\end{align}

It is now possible to show that the three function series,
\begin{align}
\label{series_b6}
&\sum_{ n=0}^{ m}\sum_{\vec c\in\{0,1\}^{\otimes { n}}}|f(\vec c)|x^{{ n}-|\vec c|}y^{|\vec c|},
\\
\label{series_b7}
&\sum_{{ n}=0}^{ m}\sum_{\vec c\in\{0,1\}^{\otimes { n}}}(n+2)|g(\vec c)|x^{{ n}-|\vec c|}y^{|\vec c|}, \quad\mbox{and}
\\
\label{series_b8}
&\sum_{{ n}=0}^{ m}\sum_{\vec c\in\{0,1\}^{\otimes { n}}}|h(\vec c)|x^{{ n}-|\vec c|}y^{|\vec c|},
\end{align}
locally converge uniformly on 
 $(x,y)\in\Omega$ as $m\rightarrow\infty$, where $\Omega:=\{(x,y)|x\geq 0,\;y\geq 0,\; x+y<\log2\}$.
The locally uniform convergence of these series implies that when $\|A\|_\mathrm{op}+\|B\|_\mathrm{op}<\log 2$, changing the order of the terms in the following series is allowed; that is, such a change has no effect on the final result: 
\begin{align}
&\sum_{n=0}^\infty\sum_{\vec c\in\{0,1\}^{\otimes n}}f(\vec c)A^{1-c_1}B^{c_1}A^{1-c_2}B^{c_2}\cdots A^{1-c_n}B^{c_n}, \label{series_b9} \\
&\sum_{n=0}^\infty\sum_{\vec c\in\{0,1\}^{\otimes n}}
(n+2) g(\vec c)A^{1-c_1}B^{c_1}A^{1-c_2}B^{c_2}\cdots A^{1-c_n}B^{c_n}, \quad \mbox{and}
\label{series_b10} \\
&\sum_{n=0}^\infty\sum_{\vec c\in\{0,1\}^{\otimes n}}h(\vec c)A^{1-c_1}B^{c_1}A^{1-c_2}B^{c_2}\cdots A^{1-c_n}B^{c_n}. \label{series_b11}
\end{align}

We now prove the locally uniform convergence of the series of (\ref{series_b6})--(\ref{series_b8}). In the following, the pair $(x_0,y_0)\in \Omega$ is chosen in $\mathbb{R}^2$ so that $0\le x\leq x_0, 0\le y\leq y_0$, and $x_0+y_0<\log 2$.

For the first series (\ref{series_b6}), we can consider the following inequalities:
\begin{align}
\infty&>
\frac{d^m}{dz^m} \left.\frac{-\log(2-e^{z})}{e^{z}-1}\right|_{ z=x_0+y_0}
\nonumber\\
&\geq
\sum_{n=0}^\infty
\frac{d^m}{dz^m} \left. a_nz^n\right|_{z=x+y} \nonumber\\
&=
\sum_{n=0}^\infty \partial_{x,y}^{(m)}
\sum_{\vec c\in\{0,1\}^{\otimes n}}f_+(\vec c)x^{n-|\vec c|}y^{|\vec c|} \nonumber\\
&=\sum_{n=0}^\infty
\sum_{\vec c\in\{0,1\}^{\otimes n}}f_+(\vec c)
\partial_{x,y}^{(m)}
x^{n-|\vec c|}y^{|\vec c|}
\nonumber\\
&\geq \sum_{n=0}^\infty\sum_{\vec c\in\{0,1\}^{\otimes n}}|f(\vec c)|
\partial_{x,y}^{(m)}
x^{n-|\vec c|}y^{|\vec c|}. 
\label{eq:infty}
\end{align}
The first inequality holds simply because the function $-\log (2-e^z)/(e^z-1)$ is analytic for $z<\log2$. The second line is a Taylor expansion of the function, where $a_n$ are all positive coefficients, and the inequality is justified because the convergence radius of the series is $\log 2$. Note that $-\log (2-e^z)/(e^z-1)=\sum_{n=1}^\infty(1/n)(e^z-1)^{n-1}$; then, $f_+(\vec{c})$ on the third line of Eq. (\ref{eq:infty}) is formally defined by the coefficients in the expansion of the following series so that
\begin{equation}\label{eq:def_fplus}
\sum_{n=1}^{\infty}\frac{1}{n}
\left(\left(\sum_{j=0}^\infty\frac{A^j}{j!}\right) \left(\sum_{j=0}^\infty\frac{B^j}{j!}\right)-I \right)^{n-1} = \sum_{n=0}^\infty\sum_{\vec c\in\{0,1\}^{\otimes n}}f_+(\vec c)A^{1-c_1}B^{c_1}A^{1-c_2}B^{c_2}\cdots A^{1-c_n}B^{c_n}
\end{equation}
holds, where $\partial_{x,y}^{(m)}$ is an $m$-th order differential operator. Specifically, this operator should be one of $d^m/dx^{m-k} dy^k$ with $0\le k\le m$ in the expansion; however, we let it represent all of these for simplicity. Additionally, $\sum_{\vec c\in\{0,1\}^{\otimes n}}f_+(\vec c) x^{n-|\vec c|}y^{|\vec c|}=a_n(x+y)^n$ is used in the third line, which can be verified through Eq. (\ref{eq:def_fplus}) by replacing $A$ and $B$ with $c$-numbers $x$ and $y$. The last inequality is due to $|f(\vec c)|<f_+(\vec c)$. 

Equation (\ref{eq:infty}) shows that when $m=0$, the series Eq. (\ref{series_b6}) converges uniformly on $(x,y)\in \Omega$. Let us define $f_\infty(x,y)$ by the RHS of Eq. (\ref{eq:infty}) in what follows.
\begin{equation}
f_\infty(x,y):=\sum_{n=0}^\infty\sum_{\vec c\in\{0,1\}^{\otimes n}}|f(\vec c)| x^{n-|\vec c|}y^{|\vec c|}.
\end{equation}

To show the convergence of the second series (\ref{series_b7}), we proceed in a similar manner:
\begin{align}
\infty&>\left.\frac\partial{\partial t}f_\infty(s,t)\right|_{s=x_0,t=y_0}
+\left.\frac\partial{\partial t}f_\infty(s,t)\right|_{s=y_0,t=x_0}
\nonumber\\
&\geq
\sum_{n=0}^\infty\sum_{\vec c\in\{0,1\}^{\otimes n}}|\vec c|\cdot
|f(\vec c)|\left(x^{n-|\vec c|}y^{|\vec c|-1}
 +y^{n-|\vec c|}x^{|\vec c|-1}\right)
\nonumber\\
&\geq
\sum_{n=1}^\infty\sum_{\vec c\in\{0,1\}^{\otimes (n-1)}}|f(c_1,c_2\cdots,c_{n-1},1)|x^{n-|\vec c|-1}y^{|\vec c|}
\nonumber\\
&\quad +
\sum_{n=1}^\infty\sum_{\vec c\in\{0,1\}^{\otimes (n-1)}}|f(1-c_1,1-c_2\cdots,1-c_{n-1},1)|x^{n-|\vec c|-1}y^{|\vec c|}
\nonumber\\
&
\geq \sum_{n=0}^\infty\sum_{\vec c\in\{0,1\}^{\otimes n}}(n+2)|g(\vec c)|x^{n-|\vec c|}y^{|\vec c|}=:g_{\infty}(x,y).
\label{eq:infty_2}
\end{align}
The first inequality comes from Eq. (\ref{eq:infty}) by setting $\partial_{x,y}^{(m)}$ to $\partial/\partial y$. The second inequality is due to the positivity of the coefficients, as well as to $0\le x\le x_0$ and $0\le y \le y_0$. The third line is obtained by multiplying each term by a number that is less than or equal to 1 and dropping some positive terms. The last line is obtained from the definition of $g(\vec{c})$ in Eq. (\ref{def:g}).

The convergence of the third series in Eq. (\ref{series_b8}) can be shown as follows:
\begin{align}
\infty&\geq
\left.\frac{d^2}{dt^2}f_\infty(s,t)\right|_{ s=x_0,t=y_0}
+\left.\frac{d^2}{dsdt}f_\infty(s,t)\right|_{ s=y_0,t=x_0}
+\left.\frac{d^2}{dsdt}f_\infty(s,t)\right|_{ s=x_0,t=y_0}
+\left.\frac{d^2}{ds^2}f_\infty(s,t)\right|_{ s=y_0,t=x_0}
+ g_\infty(x_0,y_0)^2
\nonumber\\
&\geq
\sum_{n=0}^\infty\sum_{\vec c\in\{0,1\}^{\otimes n}}
|f(\vec c)|\left(|\vec c|(|\vec c|-1)x^{n-|\vec c|}y^{|\vec c|-2}+
(n-|\vec c|)|\vec c|y^{n-|\vec c|-1}x^{|\vec c|-1}\right)
\nonumber\\
&\quad +
\sum_{n=0}^\infty\sum_{\vec c\in\{0,1\}^{\otimes n}}
|f(\vec c)|\left((n-|\vec c|)|\vec c|x^{n-|\vec c|-1}y^{|\vec c|-1}+
(n-|\vec c|)(n-|\vec c|-1)y^{n-|\vec c|-2}x^{|\vec c|}\right)
\nonumber\\
&\quad +
\left(
\sum_{n=0}^\infty\sum_{\vec c\in\{0,1\}^{\otimes n}}
(n+2)|g(\vec c)|x^{n-|\vec c|}y^{|\vec c|}\right)^2
\nonumber\\
&\geq
\sum_{n=0}^\infty\sum_{\vec c\in\{0,1\}^{\otimes n}}
\frac2{n+2}
\left(|f(1,c_1,c_2\cdots,c_{n},1)|
+
|f(0,1-c_1,1-c_2\cdots,1-c_{n},1)|\right)
x^{n-|\vec c|}y^{|\vec c|}
\nonumber\\
&\quad +
\sum_{n=0}^\infty\sum_{\vec c\in\{0,1\}^{\otimes n}}
\frac2{n+2}
\left(|f(0,c_1,c_2\cdots,c_{n},1)|
+
|f(1,1-c_1,1-c_2\cdots,1-c_{n},1)|\right)
x^{n-|\vec c|}y^{|\vec c|}
\nonumber\\
&\quad +
\left(
\sum_{n=0}^\infty\sum_{\vec c\in\{0,1\}^{\otimes n}}
|g(\vec c)|x^{n-|\vec c|}y^{|\vec c|}\right)^2
\nonumber\\
&\geq
\sum_{n=0}^\infty\sum_{\vec c\in\{0,1\}^{\otimes n}}2|g(1,c_1,c_2\cdots,c_{n})|x^{n-|\vec c|}y^{|\vec c|}
+
\sum_{n=0}^\infty\sum_{\vec c\in\{0,1\}^{\otimes n}}2|g(0,c_1,c_2\cdots,c_{n})|x^{n-|\vec c|}y^{|\vec c|}
\nonumber\\
&\quad +
\left(
\sum_{n=0}^\infty\sum_{\vec c\in\{0,1\}^{\otimes n}}
|g(\vec c)|x^{n-|\vec c|}y^{|\vec c|}\right)^2
\nonumber\\
&\geq
\sum_{n=0}^\infty\sum_{\vec c\in\{0,1\}^{\otimes n}}|h(\vec c)| x^{n-|\vec c|}y^{|\vec c|}.
\end{align}
Here, $\partial_{x,y}^{(m)}$ in Eq. (\ref{eq:infty}) is replaced with $d^2/dx^2, d^2/dxdy$, and $d^2/dy^2$, and the remaining changes are almost the same as those in Eq. (\ref{eq:infty_2}). 

Now, we are ready to define $M(\cdot,\cdot)$. In the following, we use the notation $\ad(A)(\cdot)=[A,\cdot]$, and the operator norm of this linear transformation is given by
\begin{equation}\label{opNormOfad}
\left\|\ad(A)\right\|_\mathrm{op}:=\sup_{B}\frac{\| [A,B]\|_F}{\|B\|_F},
\end{equation}
where the supremum is taken over all complex matrices $B$ whose Frobenius norms are nonzero, $\|B\|_F\neq 0$.

\begin{definition}\label{def:M_appE}
When $\|\ad(A)\|_\mathrm{op}+\|\ad(B)\|_\mathrm{op}<\log2$ holds,
$M(A,B):Hom(\mathbb C^N)\times Hom(\mathbb C^N)\rightarrow Hom(\mathbb C^N)$ is defined as follows:
\begin{align}
M(A,B) &:=A+B+\sum_{n=0}^\infty\sum_{\vec c\in\{0,1\}^{\otimes n}}g(\vec c)\ad(A,B)^{\vec c}([A,B])
\nonumber\\
&=A+B+\frac12[A,B]+\frac1{12}\left([A,[A,B]]+[B,[B,A]]\right) +\cdots,
\label{def:M}
\end{align}
where for $\vec c=(c_1,c_2,\cdots,c_n)$,
\begin{align}
\ad(A,B)^{\vec c}&:=\ad(A)^{1-c_1} \circ \ad(B)^{c_1}\circ \ad(A)^{1-c_2} \circ \ad(B)^{c_2}\circ \cdots\circ  \ad(A)^{1-c_n}\circ  \ad(B)^{c_n}.
\end{align}
\end{definition}
For example, if $\vec c=(0,1,1)$, then $\ad(A,B)^{\vec c}=\ad(A)\circ \ad(B)\circ \ad(B)$. 
$\ad(A,B)^{\varnothing}$ is considered to be identity as a superoperator.

We can now obtain a relation between the norms of operators.
\begin{lemma}
\label{lem:norm}
For any linear operator $A$, the following relation holds:
\begin{align}
\left\|\ad(A)\right\|_\mathrm{op} \leq 2\left\|A\right\|_\mathrm{op} \leq 2\left\|A\right\|_F.
\end{align}
\end{lemma}

Note that when $A, B$ are both anti-Hermitian, $M(A,B)$ is also anti-Hermitian. When $A, B$ are Hermitian, $iM(iA,iB)\in \mathcal{L}(\{iA,iB\})$. 
Although $h(\vec{c})$ defined in Eq. (\ref{def:h}) is unnecessary for $M(\cdot,\cdot)$, we define it because it will be used in the proof of Eq. (\ref{eq:abs_rel_1}).

\section{Proof of Eq. (\ref{eq:abs_rel_1})} \label{app:proof_Minequality}
We now prove Eq. (\ref{eq:abs_rel_1}), i.e., 
\begin{equation}\label{M_appF}
\|M(A,B)\|_F^2\leq 2 \|A\|_F^2+2\|B\|_F^2-\|A-B\|_F^2.\tag{\ref{eq:abs_rel_1}}
\end{equation}

\begin{lemma}
\label{lem:normFor46}
When $A$ and $B$ are anti-Hermitian and $\|A\|_F,\|B\|_F< \Delta$, Eq. (\ref{eq:abs_rel_1})
holds, where
\begin{align}
\Delta &:=\min\left(
\frac{\log2}6\delta, \frac{1}{4}\log 2\right),
\label{def:delta}
\\
\delta &:=\min\left(\frac{1}{12}\left(\sum_{n=1}^\infty\sum_{\vec c\in\{0,1\}^{\otimes n}}|h(\vec c)|\left(\frac{\log2}3\right)^n\right)^{-1}, 1\right)>0.
\end{align}
\end{lemma}
Note that since $2\times \log 2/3 <\log 2$, the infinite series in the definition of $\delta$ converges. 
Additionally, from Lemma \ref{lem:norm}, we can see that the condition for defining $M(A,B)$ is met (see Def. \ref{def:M_appE}):
\begin{align}
\|\ad(A)\|_\mathrm{op}+\|\ad(B)\|_\mathrm{op} \leq 2\|A\|_\mathrm{op}+2\|B\|_\mathrm{op} \leq 2\|A\|_F+2\|B\|_F <4\Delta\leq \log 2.
\end{align}

\begin{proof}
The difference between the left- and right-hand sides of Eq. (\ref{eq:abs_rel_1}) is
\begin{align}
&
-\|M(A,B)\|_F^2+ 2\|A\|_F^2+2\|B\|_F^2 - \|A-B\|_F^2
\nonumber\\
&= \mathrm{tr}M(A,B)^2-2\mathrm{tr}A^2-2\mathrm{tr}B^2+\mathrm{tr}(A-B)^2
\nonumber\\
&=
\sum_{n=0}^\infty\sum_{\vec c\in\{0,1\}^{\otimes n}}2g(\vec c){\rm tr}(A+B)\ad(A,B)^{\vec c}([A,B])
\nonumber\\
&\quad+
\sum_{n=0}^\infty \sum_{m=0}^\infty
\sum_{\vec c\in\{0,1\}^{\otimes n}}
\sum_{\vec c'\in\{0,1\}^{\otimes m}}
g(\vec c) g(\vec c^\prime)
{\rm tr}\left(\ad(A,B)^{\vec c}([A,B])\ad(A,B)^{\vec c'}([A,B])\right)
\nonumber\\
&=
\sum_{n=1}^\infty\sum_{\vec c\in\{0,1\}^{\otimes n-1}}2g(1,c_1,c_2,\cdots ,c_{n-1}){\rm tr}\left([A,B]\ad(A,B)^{\vec c}([A,B])\right)
\nonumber\\
&\quad-
\sum_{n=1}^\infty\sum_{\vec c\in\{0,1\}^{\otimes n-1}}2g(0,c_1,c_2,\cdots ,c_{n-1}){\rm tr}\left([A,B]\ad(A,B)^{\vec c}([A,B])\right)
\nonumber\\
&\quad+
\sum_{n=0}^\infty \sum_{m=0}^\infty
\sum_{\vec c\in\{0,1\}^{\otimes n}}
\sum_{\vec c'\in\{0,1\}^{\otimes m}}
(-1)^n g(\vec c) g(\vec c')
{\rm tr}\left([A,B]\ad(A,B)^{(c_n,\cdots,c_2,c_1,c_1',c_2'\cdots,c_m')}([A,B])\right)
\nonumber\\
&=
{\rm tr}\left([A,B]
\left(
\sum_{n=0}^\infty\sum_{\vec c\in\{0,1\}^{\otimes n}}h(\vec c)\ad(A,B)^{\vec c}
\right)([A,B])\right).
\label{eq:ap_3_1}
\end{align}
The second equality is a result of substituting $M$ in Eq. (\ref{def:M}), and the following four relations (\ref{Ff6}--\ref{Ff9}) are used to rewrite the first term of the third line; Eq. (\ref{Ff10}) is used for the second term:
\begin{align}
\mathrm{tr}\left(A \ad(A,B)^{(0,c_2,c_3,\cdots,c_n)}(C)\right) &=0,
\label{Ff6}
\\
\mathrm{tr}\left(B \ad(A,B)^{(1,c_2,c_3,\cdots,c_n)}(C)\right) &=0,
\label{Ff7}
\\
\mathrm{tr}\left(A \ad(A,B)^{(1,c_2,c_3,\cdots,c_n)}(C)\right) &= {\rm tr}\left([A,B] \ad(A,B)^{(c_2,c_3,\cdots,c_n)}(C)\right),
\label{Ff8}
\\
\mathrm{tr}\left(B \ad(A,B)^{(0,c_2,c_3,\cdots,c_n)}(C)\right) &= -\mathrm{tr}\left([A,B] \ad(A,B)^{(c_2,c_3,\cdots,c_n)}(C)\right),
\label{Ff9}
\\
\mathrm{tr}\left(\ad(A,B)^{(c_1,c_2,\cdots, c_n)}(C)\circ \ad(A,B)^{(c_1',c_2',\cdots,c_m')}(C)\right) &=
(-1)^n \mathrm{tr}\left(C \ad(A,B)^{(c_n,\cdots, c_2,c_1,c_1',c_2',\cdots,c_m')}(C)\right).
\label{Ff10}
\end{align}
The definition of $h(\vec{c})$, Eq. (\ref{def:h}), is used directly to obtain the last equality in Eq. (\ref{eq:ap_3_1}). The order of summation is changed for all equalities in Eq. (\ref{eq:ap_3_1}), but this does not affect the result. 

Furthermore, the last expression in Eq. (\ref{eq:ap_3_1}) can be evaluated by using the explicit value $h(\varnothing)=-1/12$, as well as simple relations such as $|\bra x X\ket x|\leq |\left<x|x\right >|\cdot \|X\|_\mathrm{op}$ and $[A,B]^\dagger=-[A,B]$, as follows:
\begin{align}
\mathrm{Eq.~(\ref{eq:ap_3_1})} &=
\frac{1}{12}\left\|\left[A,B\right]\right\|_F^2
+{\rm tr}\left([A,B]\left(\sum_{n=1}^\infty\sum_{\vec c\in\{0,1\}^{\otimes n}}h(\vec c)\ad(A,B)^{\vec c}\right)([A,B])\right)
\nonumber\\
&\geq
\frac1{12}\left\| \left[A,B\right] \right\|_F^2
-\left|{\rm tr}\left( [A,B]\left(\sum_{n=1}^\infty\sum_{\vec c\in\{0,1\}^{\otimes n}}h(\vec c)\ad(A,B)^{\vec c}\right)([A,B])\right) \right|
\nonumber\\
&\geq
\left\| \left[A,B\right] \right\|_F^2\left(\frac1{12}
-\left\|\sum_{n=1}^\infty\sum_{\vec c\in\{0,1\}^{\otimes n}}h(\vec c)\ad(A,B)^{\vec c}\right\|_\mathrm{op}\right)
\nonumber\\
&\geq
\left\| \left[A,B\right] \right\|_F^2\left(\frac1{12}
-\sum_{n=1}^\infty\sum_{\vec c\in\{0,1\}^{\otimes n}}|h(\vec c)|\left\|\ad(A)\right\|_\mathrm{op}^{n-|\vec c|}\left\|\ad(B)\right\|_\mathrm{op}^{|\vec c|}
\right)
\nonumber\\
&\geq
\left\| \left[A,B\right]\right\|_F^2\left(\frac1{12}
-\delta\sum_{n=1}^\infty\sum_{\vec c\in\{0,1\}^{\otimes n}}|h(\vec c)|
\left(\delta^{-1}\left\|\ad(A)\right\|_\mathrm{op}\right)^{n-|\vec c|}\left(\delta^{-1}\left\|\ad(B)\right\|_\mathrm{op}\right)^{|\vec c|}
\right)
\nonumber\\
&\geq
\left\| \left[A,B\right] \right\|_F^2\left(\frac1{12}
-\delta\sum_{n=1}^\infty\sum_{\vec c\in\{0,1\}^{\otimes n}}|h(\vec c)|
\left(\frac{\log2}3\right)^{n}
\right) \\
&\geq 0.
\end{align}
To obtain the fourth relation, we use the subadditivity and submultiplicativity of the operator norm. 
The fifth relation holds because every term in the summation has at least one of the factors
$\left\|\ad(A)\right\|_\mathrm{op}$ and $\left\|\ad(B)\right\|_\mathrm{op}$.
The second-to-last inequality is due to  
\begin{align}
\left\|\ad(A)\right\|_\mathrm{op} &\leq 2\left\|A\right\|_\mathrm{op}
\leq 2\left\|A\right\|_F
\leq 2\Delta\leq\frac{\log2}3\delta,
\end{align}
which can be derived directly from the assumption and Lemma \ref{lem:norm}. The last inequality comes from the definition of $\delta$. Hence, Eq. (\ref{eq:ap_3_1}) is shown to be positive; thus, Eq. (\ref{eq:abs_rel_1}) is proven.
\end{proof}

\section{Proof of Lemma \ref{lem:bch_dev_triangle}}
\label{m_app:proof_Lemma1}

This section describes the proof of Lemma \ref{lem:bch_dev_triangle}. In what follows, we let $\tilde{X}:= X-(\tr X)/D\cdot I$ be a traceless matrix for any $D\times D$ matrix $X$. Also we shall often use a concise relation for such traceless matrices $\tilde{X}$,
\begin{equation}\label{eq:traceless_dev_Fnorm}
\dev \tilde{X} =\frac{1}{\sqrt{D}} \|\tilde{X}\|_F,
\end{equation}
which is from the definition of $\dev(\cdot)$ in Eq. (\ref{eq:dev_definition}).

\begin{proof}
Let $m_a$ and $m_b$ be the numbers of divisions of time during which operators $A$ and $B$ are applied, respectively, and define their values as
\begin{align}\label{m_def:ma}
m_a &:=\left\lceil \Delta^{-1}r \sqrt{D}\dev A\right\rceil,
\\
\label{m_def:mb}
m_b &:=\left\lceil \Delta^{-1} r \sqrt{D} \dev B\right\rceil.
\end{align}
The total number of divisions is
\begin{equation}\label{m_def:n}
n=m_a+m_b.
\end{equation}
In Eqs. (\ref{m_def:ma}) and (\ref{m_def:mb}), $\Delta$ is the upper bound of the operator norms defined in Eq. (\ref{def:delta}), namely,
\begin{equation}\label{def_Delta}
\Delta =\min\left(\frac{\log2}6\delta, \frac{1}{4}\log 2\right).
\end{equation}
$r$ is a number that roughly parameterizes the degree of division and is set as
\begin{align}
r\in \{\mathbb{N}_{>1} |\; m_a \;\mbox{ is odd.}\}.
\label{m_def:r}
\end{align}
The value of $r$ is chosen so that $m_a$ is odd for convenience. The larger $r$, the finer the divisions.
Then, the operators are divided into $n$ pieces, and sequences of operators $C_j^{(k)}$ are defined by Eqs. (\ref{def:Cjk}) and (\ref{def:Cjkplus1}). Note that $C_j^{(k)}$ are in $\mathcal{L}$ by definition.

Before going into the proof of Lemma \ref{lem:bch_dev_triangle}, let us show 
\begin{align}
\dev C_j^{(k)} &< \frac{\Delta}{\sqrt{D}}
\label{m_eq:th_main_cond_2}
\end{align}
by induction. 

If, for a given $j$, we assume that Eq. (\ref{m_eq:th_main_cond_2}) holds for any $k$, we have $\|\ad(C_j^{(k)})\|_\mathrm{op} < \log2 /2$. This is becasuse
\[
\|\ad(C_j^{(k)})\|_\mathrm{op}= \|\ad(\tilde{C}_j^{(k)})\|_\mathrm{op} < 2\|\tilde{C}_j^{(k)}\|_F = 2\sqrt{D}\dev C_j^{(k)} <\Delta < \frac{\log 2}{2},
\]
which can be shown by using Eq. (\ref{opNormOfad}), Lemma \ref{lem:norm}, Eq. (\ref{eq:traceless_dev_Fnorm}), and Eq. (\ref{def_Delta}). Thus $C_{j+1}^{(k)}$ can be defined by Definition \ref{def:M_appE} and Eq. (\ref{def:Cjkplus1}). 
%$|\rm ad C_j^{(k)}|_{op}=|\rm ad\tilde C_j^{(k)}|_{op}<2|\tilde C_j^{(k)}|_{F}=2\sqrt D dev C_j^{(k)}<\Delta<\log2/2$
%一つ目の等式はEq. (\ref{opNormOfad})から二つ目の不等式はlemma5から３つ目の関係式が定義から。4つ目の関係式が仮定から最後の不等式はΔの定義から保証される。

When $k=1$, 
\begin{align}
\dev C_j^{(1)} \leq \max\left(\frac{\dev A }{\left\lceil \Delta^{-1}r \sqrt{D} \dev A \right\rceil},\frac{\dev B }{\left\lceil \Delta^{-1}r \sqrt{D} \dev B \right\rceil}\right)\leq \frac{\Delta}{r\sqrt{D}}< \frac{\Delta}{\sqrt{D}}.
\end{align}
Assuming Eq. (\ref{m_eq:th_main_cond_2}) that $\dev C_j^{(k)} < \Delta/\sqrt{D}$, we can evaluate $\dev  C_j^{(k+1)}$ as follows:
\begin{align}
\dev C_1^{(k+1)}&\leq \max\left(
\frac 12\dev M(C_1^{(k)},C_{2}^{(k)}),
\dev C_1^{(k)}
\right)
< \frac{\Delta}{\sqrt{D}},
\label{m_appG8}
\\
\dev C_n^{(k+1)}&\leq \max\left(
\frac 12\dev M(C_{n-1}^{(k)},C_{n}^{(k)}),
\dev C_n^{(k)}
\right)
< \frac{\Delta}{\sqrt{D}},
\label{m_appG9}
\\
\dev C_j^{(k+1)}&\leq \max\left(
\frac 12\dev M(C_j^{(k)},C_{j+1}^{(k)}),
\frac 12\dev M(C_{j-1}^{(k)},C_{j}^{(k)})
\right)
< \frac{\Delta}{\sqrt{D}},
\label{m_appG10}
\end{align}
for $1<j<n$ and $k>0$.
The inequalities on the left sides of Eqs. (\ref{m_appG8})--(\ref{m_appG10}) are simply due to the definition of $C_j^{(k)}$, and those on the right are obtained via Lemma \ref{lem:normFor46}: If the anti-Hermitian operators $X,Y$ satisfy $\dev X,\dev Y < \Delta/\sqrt{D}$, then
\begin{align}
\left(\frac 12\dev M(X,Y)\right)^2 &= 
\frac{1}{D}\left(\frac{1}{2}\left\|M(\tilde X,\tilde Y)\right\|_F \right)^2 \nonumber \\
&\leq
 \frac{1}{2D} \| \tilde X\|_F^2+
 \frac{1}{2D} \| \tilde Y\|_F^2-\frac{1}{4D} \|\tilde X-\tilde Y\|_F^2 \nonumber\\
&=
 \frac 12 (\dev  X)^2+
 \frac12  (\dev Y)^2-\frac{1}{4D} \|\tilde X-\tilde Y\|_F^2 \nonumber \\
& < \frac{\Delta^2}{D}-\frac{1}{4D} \|\tilde X-\tilde Y \|_F^2 \nonumber \\
& \leq \frac{\Delta^2}{D}.
\end{align}
In the first equality, we have used $\dev M(A,B)=\dev M(\tilde{A},\tilde{B})$, which can be verified by $M(A,B)=M(\tilde{A},\tilde{B})+(\tr (A+B)/D)I$. The remaining (in)equalities are obtained by Lemma \ref{lem:normFor46}, Eqs. (\ref{def_Delta}) and (\ref{eq:traceless_dev_Fnorm}), and the assumption for induction. 
%一つ目の等式は、$dev X=dev \tilde X$と、$dev \tilde{M(A,B)}=dev M(\tilde A,\tilde B)$から得られる。ただし、後者と等式は、$M(X,Y)=M(\tilde X,\tilde Y)+({\rm tr}(X+Y)/D)I$から確認できる。二つ目の関係式は、Lemma6と、Δの定義と数学的帰納法の仮定から保証される。３つ目の等式は、定義から保証される。４つ目の関係式は数学的帰納法の仮定はら保証される。最後の不等式は自明。

Therefore, inequality (\ref{m_eq:th_main_cond_2}) is proven.
Figure 2 in the main text shows how the sequence of operators $\|C_j^{(k)}\|_F$ evolves via repeated applications of the BCH formula (Theorem \ref{th:BCH}).

Now it can be seen why we set $m_a$ as odd. The reason is to ensure that $\exp(A/m_a)$ and $\exp(B/m_b)$ at the border can be merged by the first application of Theorem \ref{th:BCH} (see Fig. \ref{fig:def_Cjk}). This is useful in avoiding a tedious division into cases depending on $k$ when we examine the convergence of $C_j^{(k)}$ in terms of $k$.

Now that the condition for Eq. (\ref{eq:abs_rel_1}), viz., Eq. (\ref{m_eq:th_main_cond_2}), is met, we can show
\begin{align}
u^{(k)}_1 - u^{(k+1)}_1&\geq \frac12 w^{(k)}_1,
\label{m_eq:th_dic_cond}
\\
u^{(k)}_2 - u^{(k+1)}_2&= \frac12 w^{(k)}_2
\label{m_eq:th_dic_cond_2}
\end{align}
for $k>0$, where $u^{(k)}_1$, $u^{(k)}_2$, $w^{(k)}_1$ and $w^{(k)}_2$ are defined 
in Eqs. (\ref{m_def:u_k})--(\ref{m_def:d_k_2}).
\if0
\begin{align}
\label{m_def:u_k}
u^{(k)}_1 &:= \sum_{j=1}^{n}\left(\dev  C_j^{(k)} \right)^2\\
\label{m_def:u_k_2}
u^{(k)}_2 &:= \sum_{j=1}^{n}\left|{\rm tr}  C_j^{(k)} \right|^2\\
\label{m_def:d_k}
d_1^{(k)} &:= \sum_{j=1}^{n-1} \left(\dev \left(C_{j}^{(k)}-C_{j+1}^{(k)}\right)\right)^2
\\
\label{m_def:d_k_2}
d_2^{(k)} &:= \sum_{j=1}^{n-1} \left|{\rm tr} \left(C_{j}^{(k)}-C_{j+1}^{(k)}\right)\right|^2
\end{align}
for $k>0$, then, as shown in Section VII of the Supplemental Material, 
$\{u^{(k)}\}$ is a nonincreasing sequence with respect to $k$, and $w^{(k)}_1$と $w^{(k)}_2$ tend toward 0 as $k$ goes to infinity; i.e.,
\begin{align}
\label{m_eq:mon_u}
& u^{(1)}_1\geq u^{(2)}_1\geq \cdots \geq 0, \;\mathrm{and} 
\\
\label{m_eq:mon_u_2}
& u^{(1)}_2\geq u^{(2)}_2\geq \cdots \geq 0, \;\mathrm{and} 
\\
\label{m_eq:van_d}
& \lim_{k\rightarrow\infty} w^{(k)}_1 =0.
\\
\label{m_eq:van_d_2}
& \lim_{k\rightarrow\infty} w^{(k)}_2 =0.
\end{align}
\fi
This inequality is important when deriving the properties of $u^{(k)}_1$, $u^{(k)}_2$, $w^{(k)}_1$ and $w^{(k)}_2$.

When $k$ is odd, using the definitions of $C_j^{(k)}, u^{(k)}_1, w^{(k)}_1, u^{(k)}_2, w^{(k)}_2$ from Eqs. (\ref{def:Cjk}), (\ref{def:Cjkplus1}), (\ref{m_def:u_k}),  (\ref{m_def:u_k_2}), (\ref{m_def:d_k}), and (\ref{m_def:d_k_2}), we have
\begin{align}\label{eq:s96}
2(u^{(k)}_1- u^{(k+1)}_1) &=
\sum_{j=1}^{\lfloor\frac n2\rfloor}
2\left(\dev C_{2j-1}^{(k)}\right)^2
+2\left(\dev C_{2j}^{(k)}\right)^2
-\left(\dev M(C_{2j-1}^{(k)},C_{2j}^{(k)})\right)^2
\nonumber\\
&\geq 
\sum_{j=1}^{\lfloor\frac n2\rfloor}
\left(\dev \left(C_{2j-1}^{(k)}-C_{2j}^{(k)}\right)\right)^2
\nonumber\\
&= w^{(k)}_1,
\end{align}
which proves Eq. (\ref{m_eq:th_dic_cond}). 
The second relation is due to Lemma \ref{lem:normFor46}, that is,
\begin{align}
(\dev M( X,Y))^2 &= \frac{1}{D}\|M(\tilde X,\tilde  Y)\|_F^2 \nonumber \\
&\leq  \frac{2}{D} \|\tilde X\|_F^2 + \frac{2}{D}\|\tilde Y\|_F^2 -  \frac{1}{D}\|\tilde X-\tilde Y\|_F^2 \nonumber \\ \\
&= 2 (\dev  X)^2+2(\dev Y)^2-\left(\dev\left( A- B\right)\right)^2
\end{align}
for which the assumption holds, i.e., Eq. (\ref{m_eq:th_main_cond_2}).
The last relation is obtained from Eq. (\ref{m_def:d_k}), noting that 
$C^{(k)}_{2j} = C^{(k)}_{2j+1}$ for $j\in\{1,2,\cdots \lfloor(n-1)/2\rfloor\}$ when $k$ is an odd number larger than $1$.
We can also show Eq. (\ref{m_eq:th_dic_cond}) for even $k$ in a similar manner. 

Equatioin (\ref{m_eq:th_dic_cond_2}) can be shown in a similar manner:
\begin{align}
2(u^{(k)}_2- u^{(k+1)}_2) &=
\sum_{j=1}^{\lfloor\frac n2\rfloor}
2\left|{\rm tr} C_{2j-1}^{(k)}\right|^2
+2\left|{\rm tr} C_{2j}^{(k)}\right|^2
-\left|{\rm tr} \left(C_{2j-1}^{(k)}+C_{2j}^{(k)}\right)\right|^2
\nonumber\\
&=
\sum_{j=1}^{\lfloor\frac n2\rfloor}
\left|{\rm tr} \left(C_{2j-1}^{(k)}-C_{2j}^{(k)}\right)\right|^2
\nonumber\\
&= w^{(k)}_2.
\end{align}
The first equality here is due to the relation derived from the definition of $M(\cdot,\cdot)$, 
\begin{equation}\label{traceM}
\tr M( X,Y)=\tr ( X+Y),
\end{equation}
and the second one holds since all matrices on both sides are anti-Hermitian, thus their traces are pure imaginary. 
The last relation is obtained from Eq. (\ref{m_def:d_k_2}).
We can also show Eq. (\ref{m_eq:th_dic_cond_2}) for even $k$ in the same way.

Because Eq. (\ref{m_eq:th_dic_cond}) holds, and $u^{(k)}_1$ and $w^{(k)}_1$ are nonnegative, $u^{(k)}_1$ is a nonincreasing sequence; thus, Eq. (\ref{m_eq:mon_u}) in the main text is obtained, which is $u^{(1)}_1\geq u^{(2)}_1\geq \cdots \geq 0$.
Additionally, Eq. (\ref{m_eq:van_d}) holds, since $\{u^{(k)}_1\}$ is a converging sequence and 
\begin{align}\label{m_eq:conv_dk}
0=\lim_{k\rightarrow\infty}2(u^{(k)}_1- u^{(k+1)}_1) \geq \lim_{k\rightarrow \infty}w^{(k)}_1\geq 0,
\end{align}
due to Eq. (\ref{eq:s96}) and the nonnegativity of $w^{(k)}_1$.

A similar relation for $u_2^{(k)}, w_2^{(k)}$,
\begin{align}\label{eq:m_conv_dk}
0=\lim_{k\rightarrow\infty}2(u^{(k)}_2 - u^{(k+1)}_2) = \lim_{k\rightarrow \infty}w^{(k)}_2\geq 0.
\end{align}
can also be obtained, so $u^{(1)}_2\geq u^{(2)}_2\geq \cdots \geq 0$ holds.

Although Eqs. (\ref{m_eq:mon_u}), (\ref{m_eq:mon_u_2}), (\ref{m_eq:van_d}) and (\ref{m_eq:van_d_2}) suggest that $\{C_j^{(k)}\}$ converges to a single operator as $k\rightarrow \infty$, it is not straightforward to prove this. Instead, here, we show the existence of an operator that has the properties required for $C$ in Lemma \ref{lem:bch_dev_triangle} by focusing on a converging subsequence $\{C_j^{(k_m)}\}_m$, which is contained in $\{C_j^{(k)}\}_k$. First, we examine the case of $j=1$, a subsequence $\{C_1^{(k_m)}\}_m$ in $\{C_1^{(k)}\}_k$. It exists because the trivial relation
$u^{(k)}_1\ge \left(\dev C_1^{(k)}\right)^2$ 
and $u^{(k)}_2\ge \left|{\rm tr}C_1^{(k)}\right|^2$ 
hold and all the elements in $\{C_1^{(k)}\}$ are in a compact subset of $\mathcal{L}(\{A,B\})$.
Now, let the operator to which the subsequence converges be $C/n\in \mathcal{L}(\{A,B\})$, that is,
\begin{align}
\label{m_def:C}
C &:= n \lim_{m\rightarrow \infty}C_1^{(k_m)},\\
k_{m_1} &< k_{m_1+m_2}, \;\;\; \forall m_1, m_2\in \mathbb{N}. 
\end{align}
Then, the subsequences of all other ``$j$-th" sequences $\{C_j^{(k_m)}\}$ converge to $C/n$ as well:
\begin{align}
\lim_{m\rightarrow \infty}C_j^{(k_m)}&=\frac{C}{n}.
\label{m_prop:conv_Cj}
\end{align}
This can be shown as follows for $0< j\leq n$
\begin{align}\label{eq:s104}
\begin{split}
\lim_{m\rightarrow \infty} \dev \left(C_j^{(k_m)}-\frac{1}{n} C\right)=0,
\\
\lim_{m\rightarrow \infty} \left|{\rm tr} \left(C_j^{(k_m)}-\frac{1}{n} C\right)\right|=0.
    \end{split}
\end{align}
The first equation of Eq. (\ref{eq:s104}) is verified as
\begin{align}
\lim_{m\rightarrow \infty} \dev \left(C_j^{(k_m)}-\frac{1}{n} C\right)
&=
\frac{1}{\sqrt{D}}\lim_{m\rightarrow \infty} \left\|\tilde C_j^{(k_m)}-\frac{1}{n} \tilde C\right\|_F
\nonumber\\
&\leq 
\frac{1}{\sqrt{D}}\lim_{m\rightarrow \infty}\left\|\tilde C_1^{(k_m)}-\frac{1}{n} \tilde C\right\|_F 
+\frac{1}{\sqrt{D}}\lim_{m\rightarrow \infty} \sum_{q=2}^{j}\left\|\tilde C_q^{(k_m)}-\tilde C_{q-1}^{(k_m)}\right\|_F
\nonumber\\
&= 
\lim_{m\rightarrow \infty} \sum_{q=2}^{j}\dev\left( C_q^{(k_m)}-C_{q-1}^{(k_m)}\right)
\nonumber\\
&\leq 
\sqrt{j-1}
\lim_{m\rightarrow \infty}\left(\sum_{q=2}^{j}\left(\dev\left( C_q^{(k_m)}-C_{q-1}^{(k_m)}\right)\right)^2\right)^{1/2}
\nonumber\\
&\leq 
\sqrt {j-1}
\lim_{m\rightarrow \infty}\sqrt{w^{(k_m)}_1} \nonumber \\
&=0.
\end{align}
The subadditivity of the Frobenius norm is used in the second line and the first term of its RHS is dropped by Eq. (\ref{m_def:C}), whereas the second term can be rewritten as the third line by using the Cauchy-Schwarz inequality.
The 5-th relation is simply due to the definition of $w^{(k)}_1$, Eq. (\ref{m_def:d_k}). 
This value is equal to zero because of Eq. (\ref{m_eq:van_d}) (or Eq. (\ref{m_eq:conv_dk})). 
The second equation of Eq. (\ref{eq:s104}) can be obtained similarly:
\begin{align}
\lim_{m\rightarrow \infty} \left|{\rm tr} \left(C_j^{(k_m)}-\frac{1}{n} C\right)\right|
&\leq 
\lim_{m\rightarrow \infty}\left|{\rm tr}\left( C_1^{(k_m)}-\frac{1}{n}  C\right)\right| +
\lim_{m\rightarrow \infty} \sum_{q=2}^{j}\left|{\rm tr}\left( C_q^{(k_m)}- C_{q-1}^{(k_m)}\right)\right|
\nonumber\\
&\leq 
\sqrt{j-1}
\lim_{m\rightarrow \infty}\left(\sum_{q=2}^{j}\left|{\rm tr}\left( C_q^{(k_m)}-C_{q-1}^{(k_m)}\right)\right|^2\right)^{1/2}
\nonumber\\
&\leq 
\sqrt {j-1}
\lim_{m\rightarrow \infty}\sqrt{w^{(k_m)}_2} \nonumber \\
&=0.
\end{align}

The proof of Lemma \ref{lem:bch_dev_triangle} will be complete once we can confirm that the operator $C$, as the accumulation point of the sequence, satisfies Eqs. (\ref{eq:lemma_product})--(\ref{eq:lemma_trace}). 

First, Eq. (\ref{eq:lemma_lie_membership}) naturally holds because of Eq. (\ref{m_prop:conv_Cj}). Second, Eq. (\ref{m_prop:conv_Cj}) implies
\begin{equation}
\exp(C) = \lim_{m\rightarrow \infty} \exp(C_1^{(k_m)}) \exp(C_2^{(k_m)}) \cdots \exp(C_n^{(k_m)}).
\label{m_expCeqexpCjs}
\end{equation}
Since for any $k\ge 0$,
\begin{align}
\exp(C_1^{(k)}) \exp(C_2^{(k)}) \cdots \exp(C_n^{(k)})
&= \exp(C_1^{(k-1)}) \exp(C_2^{(k-1)}) \cdots \exp(C_n^{(k-1)}) \nonumber \\
&\vdots \nonumber \\
&= \exp(C_1^{(1)}) \exp(C_2^{(1)}) \cdots \exp(C_n^{(1)}) \nonumber \\
&= \exp(A) \exp(B),
\label{m_th3s}
\end{align}
Eq. (\ref{m_expCeqexpCjs}) becomes 
\begin{equation}
e^C=e^A e^B,
\end{equation}
which is Eq. (\ref{eq:lemma_product}) in the main text. In Eq. (\ref{m_th3s}), we repeatedly apply Theorem \ref{th:BCH}; this corresponds to going up from a lower row to the top one in Fig. 2. The norm condition for Theorem \ref{th:BCH} is satisfied, by Eq. (\ref{m_eq:th_main_cond_2}).

It is more difficult than it may seem to show that Eq. (\ref{eq:lemma_dev_triangle}) holds. An $r$-dependence remains in a straightforward evaluation of $\dev C_j^{(k_m)}$, as shown in Eq. (\ref{m_eq:evl_Cr}) below. Therefore, to avoid this technical difficulty regarding the fineness $r$ of operator divisions, we consider the limit of $r\rightarrow \infty$, viz., $n\rightarrow\infty$ (see Eq. (\ref{m_def:r}) for the meaning of $r$). 
Assuming that $r$ is a variable, let us rewrite $C$ in Eq. (\ref{m_prop:conv_Cj}) as $C_r$ to make the dependence on $r$ explicit. Then, by considering a sequence $\{r_m\}$, where $r_m$ are selected from the set $\{r\}$ in Eq. (\ref{m_def:r}), we let $C_\infty$ be the accumulation point of $\{C_{r_m}\}$ and show that $C_\infty$ satisfies the properties required for $C$ in Lemma \ref{lem:bch_dev_triangle}.

The upper bound of $\dev C_r$ can be obtained, starting with Eq. (\ref{m_prop:conv_Cj}), as follows:
\begin{align}
\left(\dev C_r\right)^2
&=n\lim_{m\rightarrow \infty}\sum_{j=1}^n\left(\dev C_j^{(k_m)}\right)^2
\nonumber\\
&=n\lim_{m\rightarrow \infty}u^{(k_m)}_1
\nonumber\\
&\leq  n u^{(1)}_1
\nonumber\\
&= 
\left(
\left\lceil \Delta^{-1}r \sqrt{D}\dev A \right\rceil
+\left\lceil \Delta^{-1}r \sqrt{D}\dev B \right\rceil
\right)
\left( \frac{\left(\dev A\right)^2}{\lceil \Delta^{-1}r \sqrt{D}\dev A\rceil}+\frac{\left(\dev B\right)^2}{\lceil \Delta^{-1}r \sqrt{D}\dev B\rceil}\right)
\nonumber\\
&\leq \left(
\Delta^{-1}r \sqrt{D} \dev A + \Delta^{-1}r \sqrt{D}\dev B +2 
\right)
\left(
\frac{\left(\dev A\right)^2}{ \Delta^{-1}r \sqrt{D}\dev A}
+\frac{\left(\dev B\right)^2}{ \Delta^{-1}r \sqrt{D}\dev B}\right)
\nonumber\\
&=
\left(\dev A + \dev B + \frac{2\Delta}{r\sqrt{D}} \right)
\left(\dev A + \dev B\right)
\nonumber\\
&\leq
\left(\dev A + \dev B +2\Delta \right)
\left(\dev A + \dev B\right).
\label{m_eq:evl_Cr}
\end{align}
The third line is due to the monotonicity of $u^{(k)}_1$, Eq. (\ref{m_eq:mon_u}), and the fourth relation is obtained from the definitions of $n$ and $u^{(1)}_1$ in Eqs. (\ref{m_def:ma})--(\ref{m_def:n}) %,\ref{def:mb} 
and (\ref{m_def:u_k}). 
The rest is obtained by simple rearrangement. 
The last inequality holds because $r$ is taken from the set Eq. (\ref{m_def:r}).

Equation (\ref{eq:lemma_trace}) can be shown as 
\begin{align}
{\rm tr} C_r
&=
\lim_{m\rightarrow \infty}
\sum_{j=1}^n {\rm tr} C_j^{(k_m)}
\nonumber\\
&=
\sum_{j=1}^n {\rm tr} C_j^{(1)}
\nonumber\\
&=
{\rm tr} \left(A+B\right),
\label{m_eq:evl_Cr_tr}
\end{align}
where we have used Eqs. (\ref{m_prop:conv_Cj}) and (\ref{traceM}). And the last line comes from the definition of $C_j^{(1)}$ in Eq. (\ref{def:Cjk}).

Our results can be summarized as follows. For any $r$ in Eq. (\ref{m_def:r}), we can find an operator $C_r$ such that 
\begin{align}
e^{C_r} &= e^{A}e^{B},
\label{m_eq:finlal_1}
\\
C_r&\in\mathcal L(\{A,B\}),
\label{m_eq:finlal_2}
\\
\dev C_r &\leq  \sqrt{
\left(\dev A + \dev B + \frac{2\Delta}{r\sqrt{D}} \right)
\left(\dev A + \dev B\right)
},
\label{m_eq:finlal_3}
\\
\tr C_r &= \tr(A+B).
\label{m_eq:finlal_4}
\end{align}
Since Eq. (\ref{m_eq:finlal_3}) implies that there exists an upper bound on $\dev C_r$ for $r>1$, by incorporating Eqs.~(\ref{m_eq:finlal_2}) (\ref{m_eq:finlal_4}), we see that $\{C_r\}$ is contained in a compact subspace of $\mathcal{L}(\{A,B\})$. 
Therefore, there exists a sequence $\{r_m\}$ such that $\{C_{r_m}\}$ converges to an operator $C_\infty\in\mathcal L(\{A,B\})$; i.e.,
\begin{align}
\label{def:C_infty}
C_\infty &:=\lim_{m\rightarrow \infty}C_{r_m}, \\
\lim_{m\rightarrow \infty}r_m&=+\infty.
\end{align}
Thus, if $C_\infty=C$, Eq.~(\ref{eq:lemma_lie_membership}) is satisfied, as are Eqs.~(\ref{eq:lemma_product}) and (\ref{eq:lemma_trace}), since 
$e^{C_\infty}=\lim_{m\rightarrow \infty}\exp(C_{r_m})=e^{A}e^{B}$ and 
${\tr}C_\infty=\lim_{m\rightarrow \infty}{\tr} C_{r_m}={\tr}(A+B)$ due to Eqs. (\ref{m_eq:finlal_1}) and (\ref{m_eq:finlal_4}). 

Finally, from Eq. (\ref{m_eq:finlal_4}), we have
\begin{align}
\dev C_\infty &= \lim_{m\rightarrow \infty}\dev C_{r_m}
\nonumber\\
&
\leq\lim_{m\rightarrow \infty} \sqrt{ \left(\dev A +\dev B +
\frac{2\Delta}{r_m \sqrt{D}} \right) \left(\dev A+\dev B\right)
}
\nonumber\\
&= \dev A + \dev B,
\end{align}
which means that Eq.~(\ref{eq:lemma_dev_triangle}) is also satisfied.
Therefore, $C_\infty$ is the operator $C$ whose existence was to be shown for Lemma \ref{lem:bch_dev_triangle}.
\end{proof}

\section{Proofs of Theorems \ref{thm:main_dev_bound} and \ref{th:main_1}}
\label{app:proof_th}
Let us first prove Theorem \ref{thm:main_dev_bound}, namely, Eqs.~(\ref{eq:theorem_target_generator})--(\ref{eq:theorem_trace}), from Lemma \ref{lem:bch_dev_triangle} following the flow shown in 
Sec. \ref{sec:OutlineOfProof} of the main text. 

\begin{proof}[Proof of Theorem 1]
Owing to the piecewise continuity of the Hermitian operator $H(t)$, and because $H(t)$ is bounded, the generated unitary operator $U(T)$ can be written as
\begin{align}
U(T)&=\lim_{\delta\searrow 0}\exp\left(-i\delta H\left(\left\lfloor\frac{T}{\delta}\right\rfloor\delta\right)\right)\cdots \exp\left(-i\delta H(2\delta)\right) \exp\left(-i\delta H(\delta)\right).
\label{prop:c_0}
\end{align}
From Lemma \ref{lem:bch_dev_triangle}, there is an operator $C_\delta[T]$ such that 
\begin{align}
\exp\left(-iC_{\delta}[T]\right) &=
\exp\left(-i\delta H\left(\left\lfloor\frac{T}{\delta}\right\rfloor\delta \right)\right) 
\cdots \exp\left(-i\delta H(2\delta )\right)
\exp\left(-i\delta H(\delta)\right),
\label{prop:c_1}
\\
iC_\delta[T]&\in \mathcal L(\{iH(t)\}_{0\le t\le T}),
\label{prop:c_2}
\\
{\rm dev} C_\delta[T] &\leq \sum_{n=1}^{\lfloor T/\delta\rfloor}\delta{\rm dev} H(n\delta),
\label{prop:c_3}
\\
{\rm tr} C_\delta[T] &
= \sum_{n=1}^{\lfloor T/\delta\rfloor}\delta{\rm tr} H(n\delta).
\label{prop:c_4}
\end{align}
Since $H(t)$ is piecewise continuous, for any $\varepsilon>0$, there exists a $\delta_0$ such that 
for any $\delta\in(0,\delta_0)$,
\begin{equation}
\begin{split}
\left| \sum_{n=1}^{\lfloor T/\delta\rfloor}\delta {\rm dev} H(n \delta)- \int_{0}^T dt {\rm dev} H(t) \right| \leq \varepsilon,
\\
\left| \sum_{n=1}^{\lfloor T/\delta\rfloor}\delta {\rm tr} H(n \delta)
-  \int_{0}^T dt {\rm tr} H(t)
\right|\leq \varepsilon.
 \label{eq:go_tmp_0}
\end{split}
\end{equation}
From Eqs. (\ref{prop:c_3}-\ref{eq:go_tmp_0}), it can be seen that, if $\delta<\delta_0$, $\dev C_\delta[T]$ and $\tr C_\delta[T]$ are upper bounded for a fixed $T$. Thus, $\|C_\delta[T]\|_F$ is also bounded, that is, $C_\delta[T]$ is included in a compact set. 
%Tを固定したとき、δが$\delta_0$よりも小さければ、$ C_\delta[T]$のトレースとdevが有界であることが保証される。さらに、dev tr フロベニウスノルムの関係から、$| C_\delta[T]|_F$も有界である。つまり、$C_\delta[T]$コンパクト集合に含まれることが言える。

Then, it is possible to choose a decreasing sequence $\{\delta_m\}$ of $\delta$ for any fixed $T$ that converges to zero so that 
\begin{align}
iC[T] &:= i\lim_{m\rightarrow \infty}C_{\delta_m}[T],
\label{def:C(t)}
\end{align}
which is naturally in $\mathcal{L}(\{iH(t)\}_{0\le t\le T})$. Therefore, from Eq. (\ref{prop:c_1}), we have
\begin{align}
\exp(-iC[T]) &=
\lim_{m\rightarrow \infty} \exp(-iC_{\delta_m}[T]) \nonumber
\\
&= \lim_{m\rightarrow \infty} \exp\left(-i\delta H\left(\left\lfloor\frac{T}{\delta_m}\right\rfloor\delta_m \right)\right) 
\cdots \exp\left(-i\delta H(2\delta_m )\right)
\exp\left(-i\delta H(\delta_m )\right)
\nonumber\\
&=U(T).
\label{eq:go_tmp_5}
\end{align}
Additionally, Eq.~(\ref{prop:c_3}) leads to
\begin{align}
{\rm dev}C[T] & =
\lim_{m\rightarrow \infty}{\rm dev}C_{\delta_m}[T]
\nonumber\\
&\leq
\lim_{m\rightarrow \infty} 
 \sum_{n=1}^{\lfloor T/\delta_m\rfloor}\delta_m {\rm dev} H(n \delta_m)
\nonumber\\
&=
 \int_{0}^T dt
 {\rm dev}H(t)+\epsilon.
 \label{eq:go_tmp_1}
\end{align}
for any $\epsilon >0$.
Similarly, Eq.~(\ref{prop:c_4}) implies
\begin{align}
\left|{\rm tr}C[T] - \int_{0}^T dt
 {\rm tr}H(t)\right|& =
\left|\lim_{m\rightarrow \infty}{\rm tr}C_{\delta_m}[T]- \int_{0}^T dt
 {\rm tr}H(t)\right|
\nonumber\\
&=
\left|\lim_{m\rightarrow \infty} 
 \sum_{n=1}^{\lfloor T/\delta_m\rfloor}\delta_m {\rm tr} H(n \delta_m)
- \int_{0}^T dt
 {\rm tr}H(t)\right|\nonumber\\
&\leq 
\epsilon
.
 \label{eq:go_tmp_2}
\end{align}
for any $\epsilon$.
Now, Eqs. (\ref{eq:theorem_target_generator})--(\ref{eq:theorem_trace}) are proven; hence, Theorem \ref{thm:main_dev_bound} holds.
\end{proof}

Let us move on to prove Theorem \ref{th:main_1}, i.e., Eqs. (\ref{eq:main_3})-(\ref{eq:main_5}), by Lemma \ref{th:lemma_1}. This can be carried out in a similar manner to that of Theorem \ref{thm:main_dev_bound}, so below we depict some points that are different from above. 

\begin{proof}[Proof of Theorem 2]
    
For a unitary $U(T)$ with discretized time, Lemma \ref{th:lemma_1} implies there exists a $\delta$-dependent operator $C_\delta [T]$, so that
\begin{align}
\|C_\delta[T]\|_F &\leq \sum_{n=1}^{\lfloor T/\delta\rfloor}\delta \|H(n\delta)\|_F
\label{prop:c_3_1}
\end{align}
holds, instead of Eqs. (\ref{prop:c_3}) and (\ref{prop:c_4}).
Further, since $H(t)$ is piecewise continuous, for any $\varepsilon>0$, there exists a $\delta_0$ such that 
\begin{align}
\left| \sum_{n=1}^{\lfloor T/\delta\rfloor}\delta \|H(n \delta)\|_F - \int_{0}^T dt\|H(t)\|_F \right| \leq \varepsilon
\end{align}
holds, instead of Eq. (\ref{eq:go_tmp_0}).
This guarantees that $C_\delta[T]$ is in a compact space for a fixed $T$. Then, it is possible to choose a decreasing and converging-to-zero sequence $\{\delta_m\}$ of $\delta$ for any fixed $T$ so that $iC[T]$ in Eq. (\ref{def:C(t)}) is in $\mathcal{L}(\{iH(t)\}_{0\le t\le T})$. Equation (\ref{eq:go_tmp_5}) then also holds for a similar reason.
%(\ref{def:C(t)})で定義される$iC[T]$が$\mathcal{L}(\{iH(t)\}_{0\le t\le T})$に含まれるように。この時、同様の理由で(\ref{eq:go_tmp_5})が成り立つ。

Equation (\ref{prop:c_3_1}) allows us to have, for any $\varepsilon>0$,
\begin{align}
\|C[T]\|_F & =
\lim_{m\rightarrow \infty}\|C_{\delta_m}[T]\|_F
\nonumber\\
&\leq
\lim_{m\rightarrow \infty} 
 \sum_{n=1}^{\lfloor T/\delta_m\rfloor}\delta_m \|H(n \delta_m)\|_F
\nonumber\\
&=
 \int_{0}^T dt\|H(t)\|_F+\varepsilon,
\end{align}
instead of Eqs. (\ref{eq:go_tmp_1}) and (\ref{eq:go_tmp_2}).
Therefore, Eqs. (\ref{eq:main_3})-(\ref{eq:main_5}) are proven; thus, Theorem \ref{th:main_1} holds.
\end{proof}

\end{document}